\documentclass[12pt]{article}
\usepackage{a4}
\usepackage{epsf}
\usepackage{amssymb}
\usepackage{cite}
\newcommand{\be}{\begin{equation}}
\newcommand{\ee}{\end{equation}}
\newcommand{\bea}{\begin{eqnarray}}
\newcommand{\eea}{\end{eqnarray}}

\begin{document}
\def\C{{\mathbb{C}}}
\def\R{{\mathbb{R}}}
\def\s{{\mathbb{S}}}
\def\T{{\mathbb{T}}}
\def\Z{{\mathbb{Z}}}
\def\W{{\mathbb{W}}}
\def\Bbb{\mathbb}
\def\BZ{\Bbb Z} \def\BR{\Bbb R}
\def\BW{\Bbb W} 
\def\BM{\Bbb M} 
\def\e{\mbox{e}}
\def\BC{\Bbb C} \def\BP{\Bbb P}
\def\CP{\BC\BP}
\begin{titlepage}
\title{On Localized Tachyon Condensation in $\BC^2/\BZ_n$ and
$\BC^3/\BZ_n$}
\author{}
\date{
Tapobrata Sarkar
\thanks{\noindent E--mail:~ tapo@ictp.trieste.it}
\thanks{\noindent Address from 13th July, 2004: Department of Physics, IIT, 
Kanpur 208016,  India.}
\vskip0.4cm
{\sl the Abdus Salam \\
International Center for Theoretical Physics,\\
Strada Costiera, 11 -- 34014 Trieste, Italy}}
\maketitle
\abstract{We study some aspects of localized tachyon condensation
on non-supersymmetric orbifolds of the form $\BC^2/\BZ_n$ and 
$\BC^3/\BZ_n$. We discuss the gauged linear sigma models for these
orbifolds. We show how several features of the decay of 
orbifolds of $\BC^3$ can be realised in terms of orbifolds of 
$\BC^2$. 
}
\end{titlepage}

\section{Introduction}\label{intro}
In recent years, the issue of tachyon condensation on unstable 
backgrounds has been studied extensively in string theory, and has led
to a number of useful insights. In open string theory, this problem
has been investigated by using various techniques, such as the boundary state
formalism and boundary string field theory, following the pioneering work 
of Sen \cite{sen}. \footnote{See \cite{hb} for some early work related 
to the subject of tachyon condensation.} Tachyon condensation in 
closed string theory was first studied by Adams, Polchinski and 
Silverstein (APS) in \cite{aps}, and subsequently in 
\cite{vafa},\cite{hkmm}, using different tools. 

Whereas open string tachyon condensation typically involves a change
in the D-brane configuration of the system, the corresponding situation 
in the case of closed strings is very different. In general,
condensation of closed string tachyons leads to a decay of the 
space-time itself. This can be difficult to study in general in the 
presence of bulk tachyons, but the situation is considerably simplified
if the tachyons are localized on defects, such as orbifolds. 
The presence of tachyons will generically break space-time supersymmetry,
but a lot of information can still be obtained by using world sheet
RG flow methods. 

The problem originally considered in \cite{aps} was string theory in
the background of $\BR^{1,7}\times \BC/\BZ_n$. This is the well known
conical singularity with a deficit angle related to $n$. In this case, 
demanding that all the tachyonic modes are localized at the orbifold
fixed point requires that $n$ is odd. It was shown in \cite{aps} that
the $\BC/\BZ_n$ orbifold decays with time, i.e the original background
decays into orbifolds of lower rank, and this process continues till 
we finally reach flat space-time. In terms of the conical geometry, 
this implies that the tip of the cone smooths out, and at the end
of the decay we are left with flat space. 

The situation is more complicated for higher dimensional singularities. 
Whereas for the $\BC/\BZ_n$ singularities, as we will discuss 
in a while, analysis of the conformal field theory (CFT) of the 
string world sheet shows that all the twisted sectors of the theory 
are tachyonic (i.e all the twist operators give rise to tachyons in 
space-time), for $\BC^2/\BZ_n$ orbifolds, certain twisted sectors
might be marginal, and the theory can decay via the excitation
of marginal deformations.
Further, the decay process for these orbifolds stop once the 
system reaches a supersymmetric configuration, unlike the 
one-dimensional case, where the unique endpoint of tachyon 
condensation is flat space-time. Also, in certain examples, one
finds that the $\BC^2/\BZ_n$ singularity may decay into a lower
dimensional orbifold of the form $\BC/\BZ_m$. Expectedly, closed
string tachyon condensation on non-supersymmetric orbifolds of the 
form $\BC^3/\BZ_n$ has an even richer structure \cite{mnp}, since 
the resolution of these singularities is not canonical, unlike 
the lower dimensional examples. It has been shown that in this case 
the endpoint of tachyon condensation may include terminal 
singularities.  

We should also mention that the physics of tachyon condensation 
has contributed importantly to 
the understanding of some deep mathematical results on the resolution of 
singularities in complex two and three-fold orbifolds. In the 
mathematics literature, there exists a beautiful 
correspondence between the K-theory of the resolution of these
orbifolds and the representation theory of the orbifolding group. 
This is known as the McKay correspondence, and it has been well known
from the study of supersymmetric orbifolds that D-branes provide a 
physical realisation of this correspondence, in terms of branes 
wrapping cycles of the resolved singularity. It turns out that 
closed string tachyon condensation provides a way to understand a 
quantum version of the McKay correspondence \cite{mm}, \cite{mpar}. 

Much of the physics of closed string tachyon condensation can be
captured by studying the worldsheet Gauged Linear Sigma Model (GLSM) of 
Witten \cite{wittenphases} with non-supersymmetric orbifold backgrounds. 
This model, which we will describe in details in the course of the 
paper is also intimately related to toric constructions of these
orbifolds \cite{ts1}.

Although much has been done in the above context, there are certain
issues that still need to be explored. For example, consider the fact 
that the world sheet CFT description of orbifolds has a product structure,
i.e various operators in higher dimensional theories can be constructed
by simply combining lower dimensional ones. We see a reflection of this 
fact in the GLSM and toric constructions. One might expect that 
the GLSM corresponding to a supersymmetric orbifold of $\BC^3$ 
can be obtained by adding an appropriate field to that corresponding
to a non-supersymmetric $\BC^2$ orbifold, and conversely, removing (i.e 
giving a vev) to certain fields in the GLSM might result in a lower 
dimensional orbifold. It is natural to ask whether this feature is 
more generic, i.e if higher dimensional orbifold theories can be thought 
of as certain combinations of lower dimensional ones. It turns out
that this is indeed the case, and certain constructions of $\BC^3$
orbifolds by a suitable combination of $\BC^2$ orbifolds has already 
been performed in \cite{cr}. Although this has been constructed from 
the point of view of obtaining a crepant resolution to complex 
three-fold orbifolds, it might nevertheless be useful for our purposes, 
i.e for understanding closed string tachyon condensation in higher 
dimensional orbifolds in terms of lower dimensional ones. 

In this paper, we set out to understand certain issues relating to the
decay of non-supersymmetric orbifolds in lines with the above discussion. 
Our aim would be to extend and
further develop certain ideas explored in \cite{vafa},\cite{mm} in 
order to gain a better understanding of such decays. Further, we would
explore the idea of studying the decay of higher dimensional
orbifolds in terms of lower dimensional ones. 

The paper is organised as follows. Section 2 is a review section, meant
to set the notations and conventions used in the remainder of the paper.
In section 3, we begin with the GLSM method of studying the fate of
non-supersymmetric orbifolds by exploring background sigma model 
metrics of these models for 
orbifolds of $\BC^2$ and $\BC^3$ (this method has been
employed in studying the decay of $\BC/\BZ_n$ in \cite{mt}). Section
4 deals with the understanding of certain field theory aspects of 
the GLSM for unstable backgrounds, using the methods of \cite{vafa}. In
section 5, we turn to toric geometry methods of studying the same, and 
then develop upon the idea of studying flows of $\BC^3/\BZ_n$ orbifolds in
terms of orbifolds of $\BC^2$. We conclude the paper with section 6, 
by summarising our results, and suggesting some future directions. 

While this paper was being completed, reference \cite{mnp} appeared, 
which has considered localized tachyon condensation on $\BC^3/\BZ_n$
orbifolds, and we have attempted to reconcile our results 
with theirs, wherever applicable. 

\section{Closed String Tachyon condensation on $\BC^r/\BZ_n$}

In this section, we review some known facts about closed string tachyon
condensation on orbifolds of the form $\BC^r/\BZ_n$, in
order to set the notation and conventions used in the rest of the paper.
For this purpose, we will mostly deal with the $\BC/\BZ_n$ and 
$\BC^2/\BZ_n$ orbifolds. We will come back to the case $r=3$ 
in details later in the paper. All the material of this section can be found 
in \cite{aps},\cite{hkmm}, \cite{vafa},\cite{mm},\cite{ts1}.  

\subsection{Review of the APS method}

String theory in the background of orbifold singularities of the form
$\BC^r/\BZ_n$ can be studied in various ways. In the sub-stringy regime, 
we expect the world volume gauge theory of D-branes probing these 
orbifolds to be an useful tool in the study. Indeed, by constructing the
world volume gauge theories on D-p branes probing these orbifolds a'la
Douglas and Moore \cite{dm}, we can get very useful insights into the
geometry of the resolutions, which in turn have descriptions in terms
of certain toric varieties for $r > 1$ \footnote{For an introduction to 
toric varieties, the reader is referred to \cite{fulton},\cite{oda}.}. 
This is an open string description; an equivalent closed string 
picture of these orbifold singularities can be obtained in terms
of the closed string world sheet super conformal field theory (SCFT) which
enjoys $N=(2,2)$ worldsheet supersymmetry. By studying the twisted
sectors of these SCFTs, we can read off the geometrical structure
of the singularity. 

Let us begin by describing string theory on 
$\BR^{7,1}\times\BC/\BZ_n$ where $\BR^{7,1}$ denotes eight 
dimensional Minkowski space labelled by
$\left(X^0,\cdots X^7\right)$, and the quotienting
group acts on the complexified direction $Z=X^8+iX^9$ by the action
\be
Z \to \omega Z
\ee
where $\omega=\e^{\frac{2\pi i}{n}}$. This orbifold action breaks space-time
supersymmetry, and introduces tachyons in all the twisted sectors of the
orbifold conformal field theory. In order to have the tachyons localized
at the fixed point of this orbifold we require $n$ to be odd for
type II string theories \cite{aps}. 
If we consider a D-p brane probe of this singularity where the brane is
located at the orbifold fixed point and the world volume directions are 
entirely in the transverse space, we can, following \cite{dm}, construct
the world volume gauge theory on the brane, by considering the action of
$\BZ_n$ on the world volume fields, and retaining those fields that are 
invariant under the action of the group. In this way, we obtain a quiver
gauge theory of the massless world volume scalars. By giving vaccum 
expectation values (vevs) to certain scalars in the theory, we can determine 
(by examining the classical potential for the scalars) those scalars 
that become massive in the process (and hence can be integrated
out), thus resulting in the quiver gauge theory of a lower rank orbifold 
\cite{aps}. The fermionic quivers can also be determined in a similar way
by studying the Yukawa terms in the gauge theory. This construction is 
valid at substringy regimes, and far from this regime, when gravity effects
become large, one has to resort to a full supergravity analysis. APS showed
that these two regimes together give a consistent picture of the conical
singularity decaying into flat space-time.

A similar analysis can be done for the complex two-dimensional orbifold, 
by considering string theory in the background of  
$\BR^{5,1}\times\BC^2/\BZ_n$, where $\BR^{5,1}$ is flat Minkowski space of
six dimensions, labelled by the coordinates $X^0,\cdots X^5$, and the 
orbifolding group acts on the complexified directions denoted by 
$Z^1=X^6+iX^7$ and $Z^2=X^8+iX^9$ as 
\be
\left(Z^1,Z^2\right) \to \left(\omega Z^1,\omega^k Z^2\right)
\label{twofolds}
\ee
where $\omega=\e^{\frac{2\pi i}{n}}$. When $k\neq \pm 1$, this action
breaks space-time supersymmetry, and the orbifold is generically denoted
by $\BC^2/\BZ_{n(k)}$. In this case, it is difficult to classify the
flows by implementing the
tachyonic vevs directly, as in the $\BC/\BZ_n$ example. The APS procedure
here is to generate vevs for the fields in the theory in such a way as to 
maintain a certain quantum symmetry of the initial orbifolding group. 
This corresponds to the turning on of an appropriate (marginal) deformation
in the CFT which breaks the other part of the group action. As in the 
$\BC/\BZ_n$ examples, one can analyse the classical substringy regime by
resorting to brane probe techniques, and study the decay of these 
singularities, via the quiver gauge theories on the world volume of the
probe branes. The gauge theory can
be seen to give rise to the toric data for the resolution of the orbifold, as
in the supersymmetric cases. 

As an example, let us consider type II string theory in the background
of the non-supersymmetric orbifold 
$\BC^2/\BZ_{8(3)}$. The action of the quotienting group is    
\be
\left(Z^1,Z^2\right) \to \left(\omega Z^1, \omega^3 Z^2\right)
\ee
where $\omega=\e^{\frac{2\pi i}{8}}$. This orbifold has seven twisted 
sectors of which the fourth twisted sector is marginal. By working out 
the discrete group action on the world-volume scalars, and retaining
those fields that are invariant under this action, we find that the 
world volume gauge theory of a D-brane probing this orbifold
has sixteen surviving components, charged under the $U(1)^8$ gauge
group. The APS procedure is then to give vevs to a subset of these fields
so as to maintain a certain quantum $U(1)^4$ symmetry, the vevs
breaking the other part of the gauge group. This corresponds to the turning
on of the marginal deformation corresponding to the fourth twisted 
sector of the orbifold CFT. These vevs can be turned
on either via the surviving components of $Z^1$ or those of $Z^2$, and
there are two distinct configurations into which the original theory
can decay into. In this case, it turns out that the decay products obtained
by turning on the said marginal deformation is \cite{aps}
\be
\BC^2/\BZ_{8(3)} \to \BC^2/\BZ_{4(1)} \oplus \BC^2/\BZ_{4(1)}
\ee
The r.h.s of the above equation corresponds to two (infinitely 
separated) orbifolds that are supersymmetric with the opposite 
supersymmetry (i.e a change of complex structure makes the theory 
supersymmetric) compared to the usual supersymmetric 
$\BC^2/\BZ_{4(-1)}$ orbifold. If we choose one particular complex
structure, these will be non-supersymmetric orbifolds that will further
decay into flat space. 
In general, when the decay products are themselves
non-supersymmetric, we can turn on further 
perturbations, marginal or tachyonic, in order to reach a final 
supersymmetric configuration, which might be flat space or a
supersymmetric orbifold of lower rank.  

Tachyon condensation in orbifolds of the form $\BC^3/\BZ_n$ can
be similarly considered, and has a richer structure than
the lower dimensional examples \cite{mnp}. The orbifold action in
this case is given by \footnote{We will reserve the index $k_i$
to denote the action of the orbifolding group in the two-fold examples,
the index $p_i$ will be used for three-fold examples}
\be
\left(Z^1,Z^2,Z^3\right) \to 
\left(\omega Z^1,\omega^{p_1}Z^2, \omega^{p_2} Z^3\right)
\label{threefolds}
\ee
where, like before, the $Z^i$s are the orbifolded directions and the 
transverse space is now four-dimensional. As explained in \cite{mnp}, the 
situation here is more subtle, since there is no canonical resolution 
of orbifolds of $\BC^3$, and the decay process depends 
crucially on the condensation of the most relevant tachyon, 
i.e the tachyon with the highest negative mass squared. We will 
come back in details to three-fold orbifolds later in the paper.

 
We now move on to the description of these orbifolds in terms of 
the world sheet CFT of closed
strings, which is intimately related to the toric geometry of the 
resolution of these singularities. 

\subsection{World Sheet CFT Analysis}

In most cases, an equivalent picture of the APS description of the decay
of unstable non-supersymmetric orbifolds can be obtained by examining the
world sheet orbifold conformal field theory of closed strings, 
which in turn is related to methods of toric geometry in 
two and three (complex) dimensions. Let us begin with the orbifold 
$\BC/\BZ_n$. We will deal with type 0 theory here, with the delocalised
untwisted sector tachyon tuned to zero. 
In the NSR formalism, we are dealing with a single world sheet
chiral superfield $\Phi$ (which corresponds to the coordinate $\BC$), 
and the orbifold action is 
$\Phi \to \omega \Phi$, with $\omega$ being the n-th root of unity. There 
are $(n-1)$ twisted sectors, and we can construct \cite{dixon},\cite{hkmm} 
the twist operators   
\be
Y_j=\sigma_{j/n}{\mbox{exp}}\left[i\left(\frac{j}{n}\right)
\left(H-{\bar H}\right)\right]
\label{twistj}
\ee
where $\sigma_{j/n}$ is the bosonic twist-j operator and $H$ and ${\bar H}$
denote the bosonised fermions. The operators in (\ref{twistj}) 
are chiral, with R-charges $\frac{j}{n}$ in the $j$th twisted sector, 
and give rise to tachyons in space-time, with masses
\cite{hkmm}
\be
\frac{\alpha'}{4}M_j^2=-\frac{1}{2}\left(1-\frac{j}{n}\right)
\ee
Note that the most tachyonic sector is, in this case, the first twisted
sector, with the least R-charge, and the absolute value of the mass squared
of the tachyon is proportional to the deficit angle of the conical
singularity which this orbifold represents. \footnote{It was
shown in \cite{sin1} that the minimal R-charge of the $(c,c)$ ring of
the worldsheet CFT increases under tachyon condensation.}

The decay of this orbifold is studied by
perturbing the worldsheet Lagrangian by the vertex operators 
(\ref{twistj}) with appropriate couplings, and studying the deformation
of the chiral ring of the CFT due to this perturbation. 

Higher (complex) dimensional orbifolds can also be studied in the same
way. Here, the geometrical structure is richer, and in particular
the CFT methods correspond nicely to tools of toric geometry used to
study the resolution of these orbifold. Consider the 
orbifold $\BC^2/\BZ_{n}$. The twisted sector chiral operators are, 
in this case, given by combining the twist fields $Y^{(1)}$ 
and $Y^{(2)}$ of (\ref{twistj}) corresponding to the two complexified 
directions as,
\be
Y_j=Y_{\frac{j}{n}}^{(1)} Y_{\{\frac{jp}{n}\}}^{(2)}
\label{twist}
\ee
where $j=1,\cdots (n-1)$ corresponds to the $(n-1)$ twisted
sectors in the theory, and $\{x\}$ denotes the fractional part of the
real number $x$ (the integral part of the real number $x$ is denoted
by $\left[x\right]$). 
Orbifolds of this form have a convenient description in terms
of toric geometry, via the Hirzebruch-Jung continued fraction. 

In general, for orbifolds of the form $\BC^2/\BZ_{n(k)}$, 
where $k \neq n-1$, the resolution of the singularity can be 
succinctly described by the Hirzebruch-Jung continued function,
\be
\frac{n}{k}=a_1-\frac{1}{a_2-\frac{1}{a_3-\frac{1}{...\frac{1}{a_r}}}}
\label{hj}
\ee
where $a_i$ are integers, $a_i \geq 2$. The number of these
integers $r$ determines the number of $\BP^1$s that need to be
blown up in order to resolve the singularity completely, and the
integers $a_i$ determine the self intersection numbers of these
$\BP^1$s. These singularities are described in toric geometry by
a data matrix that consists of a set of points in a two dimensional
$SL(2,\BZ)$ lattice, and is of the form 
{\small
\begin{eqnarray}
{\cal T}=
\pmatrix{v_0,v_1&v_2&v_3&...&v_{r+1}\cr
}\label{data2}\end{eqnarray}}
where the interior vectors $v_i,~i=1,\cdots,r$ satisfy the relation
\be
a_iv_i=v_{i-1}+v_{i+1} 
\label{hjvectors}
\ee
where $v_0=(0, -1),~v_{r+1} = (n,-k)$, and
the $a_i$s are determined from the continued fraction (\ref{hj}). 
Equivalently, the toric data can be constructed following the 
prescription of \cite{asp}.
 
Let us first examine the supersymmetric case, i.e when $(1+p)=0
({\mbox{mod}}~n)$. We can consider either type 0 or type II  
theories. From the above discussion, we see that these 
orbifolds, which are of the type $\BC^2/\BZ_{n(-1)}$, are described in
toric geometry by a data matrix given by
{\small
\begin{eqnarray}
{\cal T}=
\pmatrix{
1&0&-1&-2&-3&...&-(n-1)\cr
0&1&2&3&4&...&n\cr
}\label{data1}\end{eqnarray}}
which corresponds to the fact that in this case, the Hirzebruch-Jung
continued fraction is $\frac{n}{(n-1)}=[2^{n-1}]$.

The kernel of this matrix specifies the $U(1)$ charges of the
chiral fields in a gauged linear sigma model that describes 
this orbifold, which we describe momentarily. Note however that 
in this case, since the toric
data is a $2 \times (n+1)$ matrix, the charge matrix will be 
$(n-1)\times n$ dimensional. Hence, the resolution of the
singularity will be described by a $U(1)^{n-1}$ GLSM. Indeed, it
is well known that a deformation of the orbifold by the $(n-1)$
marginal twist operators resolves this singularity into the 
$A_{n-1}$ ALE manifold. Decay of these orbifolds can be studied
in the way mentioned before, namely, we can perturb the world sheet
Lagrangian by chiral operators (all of which are marginal in this
case). This amounts to blowing up certain $\BP^1$s in the 
Hirzebruch-Jung geometry to infinite size \cite{mm}. 

Next, we examine non-supersymmetric orbifolds. 
Let us mention at the outset that the analysis of localised tachyon
condensation for non-supersymmetric orbifolds is different for 
type 0 and type II theories. 
This is related to the fact that chiral GSO projections of the latter can
remove some of the generators of the chiral ring of the CFT 
\cite{mm}. This implies that the resolution of the singularity cannot
be described entirely by K\"ahler deformations. These issues have
been considered in details in \cite{hkmm},\cite{mm}.

Here, the toric data matrix can be calculated from equations
({\ref{hj}) and ({\ref{hjvectors}). For example, for the 
orbifold $\BC^2/\BZ_{8(3)}$ that we have considered in the previous
subsection, the continued fraction is $\frac{8}{3}=\left[3,3\right]$,
and the toric data is given by the matrix
\be
{\cal T}~=~
\pmatrix{ 1 & 0 & -1 & -3 \cr 0 & 1 & 3 & 8 \cr  }
\label{eightthree}
\ee
This corresponds to the fact that in order to completely 
resolve the singularity of the orbifold $\BC^2/\BZ_{8(3)}$, 
one has to blow up two $\BP^1$s, with self intersection 
numbers $(-3,-3)$. In general, if the Hirzebruch-Jung continued
fraction contains $r$ integers $a_1,\cdots a_r$, implying that
a total of $r$ $\BP^1$s have to be blown up in order to resolve
the singularity completely, the toric data matrix will have
dimensions $2 \times (r+2)$, and this means that the corresponding 
GLSM will have the gauge group $U(1)^r$. 

There is an apparent puzzle here. The
number of $\BP^1$s needed to completely resolve this singularity
appears to be less than the rank $n$ of the compactly supported K-theory
lattice of the resolved space, and hence there is a mismatch
between the number of D-brane charges one gets on the resolution
and the number of independent brane charges of the orbifold.
These missing charges can be located on the Coulomb branch of
the GLSM, as shown in \cite{mm}, as opposed to the case of
supersymmetric orbifolds, where all the D-brane charges are
located in the Higgs branch of the GLSM. 

The toric data for the resolution of an orbifold singularity can also
be obtained by the brane probe procedure for type II 
backgrounds. Consider, for example, 
a D-p brane probing an orbifold of the form $\BC^2/\BZ_{n(k)}$. We can
construct the gauge theory living on the world volume of the probe brane
(which is obtained by orbifolding the D-brane gauge theory on flat space)
by following the well known procedure advocated in \cite{dm},\cite{dgm}.
The gauge theory, which is completely specified by its matter content 
(D-terms) and interactions (F-terms) can be used to extract information
about the resolution of the orbifold singularity that the brane probes. 
This is best illustrated by an example.  

Consider D branes background of the non-supersymmetric orbifold 
$\BR^{5,1}\times\BC^2/\BZ_{8(3)}$ for which the toric data is given in
eq. (\ref{data1}). The low energy theory is an orbifold of the
D-brane gauge theory in flat space and is obtained by the action of the
quotienting group on the coordinates and the Chan Paton indices. 
It is described by the analogues of the usual D and F terms as 
in the supersymmetric cases studied in \cite{dm},\cite{dgm}. 
The D-terms contain information about the charges of the $16$ surviving fields 
under the gauge group, which is $U(1)^8$ in this case, and let us denote
it by the matrix $\Delta$. 

As in the supersymmetric case, the F-term constraints, of the form 
$\left[Z^1,Z^2\right]=0$
are not all independent, and can be solved in terms of $9$ independent
fields, which we denote by $v_j,~j=1,\cdots 9$. 
The linear dependence of the variables in the F-term constraints in terms
of the $v_j$ can be  can be encoded as a matrix $K$, such that
$X_i=\prod_j v_j^{K_{ij}}$, where we have denoted the surviving components
of the $Z$ fields collectively by $X_i,~i=1\cdots 16$. 
The procedure of \cite{dgm} is to calculate the 
dual of this matrix (to avoid possible singularities) by introducing 
a set of new variables, which is a matrix $T$ satisfying ${\vec K}.{\vec T}
\geq 0$. The dual matrix defines a new set of fields $p_{\alpha}$, such that
$v_j=\prod_{\alpha}p_{\alpha}^{T_{j\alpha}}$. Typically, the number of new
fields $p_{\alpha}$ is greater than the number of independent fields 
$v_j$, hence there are new redundancies, which have to be taken care of
by introducing an additional set of $U(1)$ (or equivalently $C^*$) actions. 

We can calculate the charges of the variables $p_{\alpha}$ under the set of
the new $U(1)$s and call the resulting matrix $V_1$. The charges of the 
$p_{\alpha}$ under the original $U(1)^8$ can also be easily determined, and 
gives rise to the matrix $V_2$. Concatenating $V_1$ and $V_2$, and 
taking its kernel gives rise to the geometrical data for the resolution of
the singularity $\BC^2/\BZ_{8(3)}$ and is given by
\be
{\cal T}~=~
\pmatrix{ 1 & 0 & -1 & -1 & -3 \cr 0 & 1 & 4 & 3 & 8 \cr  }
\label{eightthreetwo}
\ee
By comparing with the toric data for the same orbifold given in 
(\ref{eightthree}), we notice that the data obtained from the
D-brane gauge theory contains an additional point, corresponding to the
marginal deformation by the fourth twisted sector, and hence describes a
non-minimal resolution of the orbifold. Eq. (\ref{eightthree}) describes
the minimal resolution corresponding to the Hirzebruch-Jung 
continued fraction $[3,3]$,
whereas eq. (\ref{eightthreetwo}) describes the non-minimal resolution
corresponding to the continued fraction $[4,1,4]$. This is actually a 
generic feature of D-brane gauge theories for branes probing 
non-supersymmetric orbifolds in two and three complex dimensions 
\cite{ts1}. 

\subsection{GLSM methods}

There is a third and very powerful method to study closed string
tachyon condensation, using the gauged linear sigma models
(GLSM) \cite{wittenphases}. Worldsheet theories that enjoy $(2,2)$ 
supersymmetry can be described in terms of the GLSM and this can be
used to study the condensation of closed string tachyons in target
space \cite{vafa}, and yield results equivalent to \cite{aps},
\cite{hkmm}. Since some of the details of the GLSM will be 
important for our future discussions, let us begin by summarising some
salient features of the GLSM. For the details, the reader is referred
to \cite{wittenphases}\cite{mp}. 

\subsubsection{The GLSM and its classical limits}

Generically, the moduli space of type II string theories compactified
on Calabi-Yau three folds contain different ``phases.'' In certain regions
of the K\"ahler moduli space, the theory might appear to be geometric, 
i.e described by a non-linear sigma model with Calabi-Yau target space,
and in other regions, the world sheet CFT appears to be non-geometric and
hence described by a Landau-Ginzburg theory. The GLSM of 
\cite{wittenphases} describes a method to study the transitions between the
various phases, as one moves around in moduli space. The idea is to 
introduce a $d=2$, $N=2$ supersymmetric field theory (obtained by 
dimensional reduction from a $d=4$, $N=1$ theory), with a certain number
of chiral and vector multiplets. One further introduces a Fayet-Iliapoulos
(F.I) D-term and a $\theta$ term. The action for a $U(1)^r$ GLSM 
is conveniently written in terms
of the chiral superfields $\Phi_i$, the twisted chiral superfields 
$\Sigma_a$, and the vector superfields $V_a$ as
\begin{eqnarray}
S&=&\int d^2zd^4\theta\left[\sum_{i=1}^{n}{\overline\Phi}_i{\mbox{exp}}
\left(2\sum_{a}Q_i^aV_a\right)\Phi_i-\sum_a\frac{1}{4e_a^2}
{\overline{\Sigma}_a}\Sigma_a\right]\nonumber\\
&~&~~~~-\sum_a\int d^2z\left(-r_aD_a+\frac{\theta_a}{2\pi i}v^a_{01}
\right)
\label{glsmaction}
\end{eqnarray}
where $a=1,\cdots,r$ denotes the $U(1)$ index, and in the second 
line $D_a$ is the Fayet-Iliapoulos D-term corresponding to the 
$a$th $U(1)$, that appears with $r_a$, the $a$th K\"ahler parameter.
Also, $v^a_{01}$ denotes a boundary electric field that can be turned
on in a two dimensional theory, with the coupling $\theta_a$. The $r_a$
and $\theta_a$ can be combined to form a complexified K\"ahler parameter.

The Calabi-Yau Landau-Ginzburg correspondence can be obtained by 
considering the classical theory in the limits $r\gg 0$ and $r \ll 0$, for
a theory with an additional superpotential interaction in 
eq. (\ref{glsmaction}) \cite{wittenphases}. Choosing 
a superpotential appropriately, the model is seen to interpolate between
a geometric phase (Calabi-Yau) and a non-geometric phase (Landau-Ginzburg).
The geometry is effectively captured by the D-term constraint
\footnote{The sign of $r$ determines the UV and IR regions of the theory.
We follow the convention of \cite{mt}, the UV is $r\gg 0$ and the 
IR is $r\ll 0$.}
\be
\sum_i Q_i^a|\phi_i|^2+r=0,~~~~\forall a
\label{dterm}
\ee
The superpotential is chosen in such a way that it is gauge invariant,
and the sum of the total $U(1)$ charges for each $U(1)$ is zero.
This latter condition, in fact, ensures that the theory enjoys 
anomaly free R-invariance. For GLSMs describing non-supersymmetric
orbifolds, this condition is violated, as we will see shortly.

For most of our discussion, we will be interested in a model without a 
superpotential. In such cases, the target space is a non-compact
orbifold. Let us illustrate this with an example, following \cite{vafa}. 
Consider a $U(1)$ GLSM, with $3$ charged fields $\phi_{-n}$, $\phi_1$ 
and $\phi_2$ with charges given by  
\be
Q=\left(-n,k_1,k_2\right)
\ee
with $n,k_1,k_2$ being positive, and $n > k_1,k_2$. 
The $U(1)$ action on the fields being given by
\be
\left(\phi_{-n},\phi_1,\phi_2\right) \to 
\left(e^{-in\theta}\phi_n,e^{ik_1\theta} \phi_1,
e^{ik_2\theta} \phi_2\right)
\label{uone}
\ee
The D-term constraint, is, in this case, from eq. (\ref{dterm})
\be
-n|\phi_{-n}|^2+k_1|\phi_1|^2+k_2|\phi_2|^2+r=0
\label{dterm1}
\ee
In the limit $r\gg 0$, for eq. (\ref{dterm1}) to be satisfied, we 
require $\phi_{-n}$ to take a large vev. In this case, the $U(1)$ is 
broken down to $\BZ_n$, as can be seen from eq. (\ref{uone}). The
fields $\phi_1$ and $\phi_2$ then transform under the $\BZ_n$ 
group as 
\be
\left(\phi_1,\phi_2\right) \to\left( 
e^{2\pi i\frac{k_1}{n}}\phi_1,e^{2\pi i\frac{k_2}{n}}\phi_2
\right)
\ee
which is the orbifold $\BC^2/\BZ_{n(k_1,k_2)}$ 
\footnote{One can set $k_1=1$ for convenience, as in this case
the Hirzebruch-Jung continued fraction can be used to read off the
geometry. This gives an equivalent theory with arbitrary $k_1$, 
with the first twisted sector of the theory with arbitrary $k_1$ 
being the $k_1$th twisted sector of the theory with $k_1=1$. Analogous
arguments hold for the $\BC/\BZ_n$ and the $\BC^3/\BZ_n$ orbifolds.} 
 
Here, we have considered the specific example of the two dimensional 
orbifold. One and three dimensional orbifolds can be 
considered in an entirely analogous fashion. 

\subsubsection{The Sigma Model Metric}

We now describe another important aspect of the GLSM which we
will use in the following sections. For a GLSM without a 
superpotential, the effective metric of the background can
be computed in the limit of infinite gauge coupling.
Starting from a $U(1)$ GLSM, if we set the gauge coupling 
$e \to \infty$, then the kinetic
term for the vector multiplet vanishes, and hence the fields
in the gauge multiplet become Lagrange multipliers. Hence, we
can now solve for the fields in the chiral multiplet in terms
of the gauge multiplet fields, and putting this back into the
kinetic term for the chiral multiplet, we can read off the
sigma model metric. Let us illustrate this with the example of
the $\BC/\BZ_n$ orbifold \cite{mt}. This can be described by
a GLSM with a single $U(1)$, with two chiral fields given by
charges $\left(-n, 1\right)$ under this $U(1)$. There is a
single Fayet-Iliopoulos parameter corresponding to the $U(1)$,
and the D-term constraint is given by
\be
|\phi_1|^2 -n|\phi_{-n}|^2 + r = 0
\ee
where the labels with the fields denote their $U(1)$ charges. 
As explained before, we can take two classical limits of this
GLSM, with $r \gg 0$ or $r \ll 0$. The solutions for
the fields will be different for these two limits. Let us
first consider the case $r \gg 0$. Here, we can solve
for the fields $\phi$ as 
\be
\phi_1 = \rho e^{i\theta_1},
~~~\phi_{-n}=\sqrt{\frac{\rho^2+r}{n}}e^{i\theta_n}
\ee
where $\rho$ and $\theta$ parametrise the (orbifolded) complex
plane. Substituting this solution into the kinetic term for the chiral
fields 
\be
L = - \int d^2z
g_{\mu\nu}D_{\alpha}\phi^{\mu}D^{\alpha}\phi^{\nu}
\label{sigmalag}
\ee
we can read off the sigma model metric $g_{\mu\nu}$,
remembering that the fields in the vector multiplet are now
solved in terms of those in the chiral multiplet. 
For this example, we can solve for the gauge fields as
\be
v_{\alpha}=\frac{\rho^2\partial_{\alpha}\theta_1 - \left(\rho^2 +
r\right)\partial_{\alpha}\theta_n}{(n+1)\rho^2 + nr}
\ee
putting this back into the Lagrangian (\ref{sigmalag}) we
get, in the limit of large $r$, the sigma model metric
\be
ds^2 = d\rho^2 + \frac{\rho^2}{n^2}d\theta^2
\ee
which corresponds to a cone with deficit angle
$2\pi\left(1-\frac{1}{n}\right)$. In the limit when $r\ll 0$,
we have a different solution for the fields $\phi$, given by
\be
\phi_n=\rho'e^{i\theta_{-n}},~~~\phi_1=\sqrt{n\rho'^2+|r|}
e^{i\theta_1}
\ee
and using the same methods as before, we obtain, in the limit 
$r\to -\infty$, the metric for flat space. This gives a nice
picture of the decay of the orbifold $\BC/\BZ_n$, in agreement
with APS \cite{mt}. 

We are now ready to address some relevant issues for tachyon 
condensation in non-supersymmetric non-compact orbifolds. We begin
with a calculation of the sigma model metric for the orbifolds
$\BC^2/\BZ_n$ and $\BC^3/\BZ_n$.

\section{Sigma Model Metrics for $\BC^2/\BZ_n$ and $\BC^3/\BZ_n$}

In this section, we compute the sigma model metric for
complex two-fold and three-fold orbifolds, using GLSM methods, 
generalising the $\BC/\BZ_n$ example of \cite{mt}. 
This is expected to give us results equivalent to those obtained
under worldsheet RG flows. 
First we consider complex two-fold orbifolds,  
with the action of the orbifolding group being given by
eq. (\ref{twofolds}). Before we proceed with the metric computation, 
let us address a few relevant issues about the closed string GLSM 
for these orbifolds, which will be useful for our discussion later. 

In general, an orbifold of the form $\BC^2/\BZ_{n(k)}$ will be
described by a GLSM with multiple $U(1)$ fields. One way to
understand this is via the toric description of these
orbifolds. As we have already mentioned, the resolution of these 
singularities, can, via methods of toric geometry be entirely 
specified by certain combinatorial data \cite{fulton},\cite{oda}. 
This data can be used to compute the charge matrix of the GLSM,
and the gauge group is $U(1)^r$, where $r$ is the number of 
integers appearing in the continued fraction expansion of
$\frac{n}{k}$. 
We must keep in mind that for the orbifolds under consideration, 
the above results can be obtained by the brane probe analysis 
(by constructing the D-brane gauge theories a'la Douglas and Moore
\cite{dm}) in the substringy regime (i.e when the expectation
value of the tachyonic perturbation is small compared to the
string scale). Far from the
substringy regime, when the $\alpha'$ corrections are large, the
brane probe picture is not very useful, and one has to revert to
a full time dependent analysis in the gravity regime. Such an
analysis turns out to be difficult, and one usually replaces time
dependence with the RG flow equations and then study the fate of
the unstable orbifold. Recently there have been some
work \cite{gh},\cite{mh} where a low energy supergravity action
has been employed with the tachyon acting as a source term for
massless fields, in order to obtain a somewhat controlled
description of the decay of non-supersymmetric orbifolds.

The computation of sigma-model metrics cannot track the decay of 
the orbifolds fully, 
but nonetheless it is instructive to study these in the classical 
limits, i.e in the UV and the IR, as has
been done in \cite{mt} for the case of the $\BC/\BZ_n$ orbifolds.
These will give results on the end points of the worldsheet RG flows,
although quantum corrections of the RG cannot be studied using
these methods. 
We consider the GLSM with a single $U(1)$ describing the 
$\BC^2/\BZ_{n(k)}$ orbifold. This has three fields charged under
the $U(1)$, the charges being $\left(-n,1,k\right)$.
Turning on a single $U(1)$ does not amount to a full resolution of the 
singularity, since we turn on only one of the various Fayet-Iliapoulos 
parameters. The D-term constraint is  
\be
|\phi_1|^2+k|\phi_k|^2-n|\phi_{-n}|^2 +r =0
\label{dterm2}
\ee
We solve for the fields as
\be
\phi_1=\rho_1e^{i\theta_1},~~\phi_k=\rho_ke^{i\theta_k},
~~\phi_{-n}= \sqrt{\frac{r+\rho_1^2+k\rho_k^2}{n}}e^{i\theta_n}
\ee
where again the labels on the components $\rho$ and $\theta$ denote, 
in an obvious way, the $U(1)$ charges associated with the fields
$\phi_i$. The kinetic term of the fields $\phi_i$ in the Lagrangian
now becomes
\begin{eqnarray}
L_{kin}&=&\left(\partial_{\alpha}\rho_1\right)^2 + 
\rho_1^2\left(\partial_{\alpha}\theta_1-v_{\alpha}\right)^2 + 
\left(\partial_{\alpha}\rho_k\right)^2+\rho_k^2
\left(\partial_{\alpha}\theta_k-kv_{\alpha}\right)^2\nonumber\\
&+&\frac{1}{n\left(r+\rho_1^2+\rho_k^2\right)}\left(\rho_1
\partial_{\alpha}\rho_1+\rho_1\partial_{\alpha}\rho_k\right)^2
\nonumber\\&+&\frac{\left(r+\rho_1^2+\rho_k^2\right)}{n}
\left(\partial_{\alpha}\theta_n + nv_{\alpha}\right)^2
\label{lkin}
\end{eqnarray}
From this, the equation of motion for the $v_{\alpha}$ can be
determined (in the limit of the gauge coupling going to infinity) 
and gives
\begin{equation}
v_{\alpha}=\frac{\rho_1^2\partial_{\alpha}\theta_1 + 
k\rho_k^2\partial_{\alpha}\theta_k-\left(r+\rho_1^2+\rho_k^2\right)
\left(\partial_{\alpha}\theta_n\right)}{(n+1)\rho_1^2 +
(nk+k^2)\rho_k^2 +nr}
\label{vmu}
\end{equation}
Putting the expression (\ref{vmu}) back in (\ref{lkin}), we can
read off the sigma model metric. The general expression is
complicated, but in the limit $r \gg 0$, which is what we will be
interested in, the metric is given by
\be
ds^2=d\rho_1^2+\frac{\rho_1^2}{n^2}d{\tilde{\theta_1}}^2 +
d\rho_k^2 + \frac{\rho_k^2}{(n/k)^2}d{\tilde{\theta_2}}^2
\label{metric1}
\ee
where ${\tilde{\theta_1}}=n\theta_1+\theta_n$ and
${\tilde{\theta_2}}=\frac{n}{k}\theta_k+\theta_n$ are gauge 
invariant variables. The geometry can be most clearly seen 
by making a gauge choice so as to fix $\phi_{-n}$ to be
real and positive, i.e setting $\theta_n=0$. This leads to 
the metric 
\be
ds^2=d\rho_1^2+\rho_1^2d\theta_1^2+d\rho_2^2+\rho_2^2d\theta_2^2
\ee
with the simultaneous identification
\footnote{Note that $\tilde{\theta_1}$ and $\tilde{\theta_2}$ both
have periodicity $2\pi$, but the metric in (\ref{metric1}) is not
that of two disjoint cones because of the identification
in (\ref{iden1}).}
\begin{eqnarray}
\theta_1\simeq \theta_1+\frac{2\pi}{n}\nonumber\\
\theta_k\simeq \theta_k+\frac{2\pi k}{n}
\label{iden1}
\end{eqnarray}

It is now natural to ask what happens in the IR, i.e when $r \to
-\infty$. In this regime, we can let either $\phi_1$ or $\phi_k$ 
to be very large. In the first case, we write the solution of the
D-term constraint (\ref{dterm2}) as 
\be
\phi_1=\sqrt{|r|+n\rho_n^2-k^2\rho_k^2}e^{i\theta_1},~~
\phi_k=\rho_ke^{i\theta_k},~~\phi_{-n}=\rho_{n}e^{i\theta_n}
\ee
Substituting the value of the gauge field 
\be
v_{\alpha}=\frac{1}{B_1}\left(A_1\partial_{\alpha}\theta_1
+k\rho_k^2\partial_{\alpha}\theta_k-n\rho_n^2\partial_{\alpha}\theta_n
\right)
\ee
where we have defined $A_1=\left(n\rho_n^2-k\rho_k^2+r\right)$ and 
$B_1=\left(A_1+k^2\rho_k^2+n^2\rho_n^2\right)$, we find that the
metric at large negative values of $r$ is the flat metric. 

If we let $\phi_k$ to be very large, and solve the D-term 
constraint (\ref{dterm2}) by
\be
\phi_1=\rho_1e^{i\theta_1},~~\phi_k =
\sqrt{\frac{|r|+n\rho_n^2-\rho_1^2}{k}}e^{i\theta_k},~~\phi_{-n} =
\rho_ne^{i\theta_n}
\ee
we find that the expression for the gauge field is now
\be
v_{\alpha}=\frac{1}{B_2}\left(\rho_1^2\partial_{\alpha}\theta_1 + 
A_2\partial_{\alpha}\theta_k-n\rho_n^2\partial_{\alpha}\theta_n\right)
\ee
where we have defined $A_2=\left(|r|+n\rho^2-\rho_1^2\right)$ and 
$B_2=\left(A_2k+n^2\rho_n^2+\rho_1^2\right)$. Again, substituting 
this value of $v_{\mu}$ in the Lagrangian, we obtain, in the
limit $r \to -\infty$, 
\be
L = \left(\partial_{\alpha}\rho_1\right)^2 +
\left(\partial_{\alpha}\rho_n\right)^2 +
\frac{\rho_1^2}{k^2}\left[\partial_{\alpha}\left(k\theta_1 - 
\theta_k\right)\right]^2 +
\frac{\rho_n^2}{k^2}\left[\partial_{\alpha}
\left(n\theta_k+k\theta_n\right)\right]^2
\ee
note that here the third term (along with the first) 
is related to a cone with a deficit 
angle $2\pi\left(1-\frac{1}{k}\right)$, in terms of the gauge
invariant variable $(k\theta_1-\theta_k)$. In the fourth term,
since $n > k$, we need to redefine the charge of $\phi_{-n}$,  
and hence, it is natural to replace $-n$ by $(2k - n)$. \footnote{Here
we are assuming that $2k > n$ (we have started with the condition
$k < n$. In general, we need to replace
$-n$ by $(pk - n)$ where $p$ is the smallest integer for 
which $(pk-n) > 0$.} Doing this
change, we get the sigma model metric in this case as
\be
ds^2=d\rho_1^2+d\rho_n^2+\frac{\rho_1^2}{k^2}
d{\tilde{\theta_1}}^2 + \frac{\rho_n^2}{\left(k/(2k-n)\right)^2}
d{\tilde{\theta_2}}^2
\ee
where now the gauge invariant variables are now
${\tilde{\theta_1}}=\left(k\theta_1-\theta_k\right)$ and
${\tilde{\theta_2}}=\left(\theta_k -
\frac{k}{(2k-n)}\theta_n\right)$

This shows that the end point of tachyon condensation is now 
a direct sum of flat space and a $\BC^2/\BZ_k$ orbifold, and from 
our previous analysis, this is equivalent to the orbifold singularity
$\BC^2/\BZ_{k(2k-n)}$. This orbifold may not be supersymmetric, and
we can turn on a second $U(1)$ to study its flow. In this way, 
we can stepwise desingularise the original singularity.    

This result, is expected, as we can see from 
eq. (\ref{dterm2}). In the limit $r \gg 0$, the field $\phi_{-n}$
takes a very large vev, and this breaks the $U(1)$ down to 
$\BZ_n$. For $r \ll 0$, one can either choose the $\phi_1$ or the
$\phi_k$ to acquire a large vev, and from the action of the $U(1)$
on these fields, it is clear that in the first case this completely
breaks the gauge group, so that we obtain the flat space metric,
and in the second case it breaks the gauge group down to $\BZ_k$.
This generalises the flow pattern of $\BC/\BZ_n$ of \cite{mt} to 
two-fold examples. 

An entirely analogous calculation can be done for the complex 
three fold orbifolds with the action given by eq. (\ref{threefolds}).
We will skip the details here, since the qualitative form of the
metric in the limits of the Fayet-Iliapoulos parameter $r \gg 0$
and $r \ll 0$ are obvious from the calculations of the two-fold
orbifolds presented above. In this case, we are dealing with four
fields $\left(\phi_{-n},\phi_1,\phi_{p_1},\phi_{p_2}\right)$. In 
the limit $r \gg 0$, the $U(1)$ is broken to $\BZ_n$ as before, and
in the opposite limit, we obtain flat space if we choose $\phi_1$ to
be very large. Choosing $\phi_{p_1}$ or $\phi_{p_2}$ to be very
large breaks the $U(1)$ to $\BZ_{p_1}$ or $\BZ_{p_2}$ respectively.

In the above, we have seen the decay of two-fold orbifolds when 
a single $U(1)$ is turned on. In principle, we can do a similar analysis
for multiple $U(1)$ gauge fields (in the supersymmetric case this  
should reduce to a version of the Eguchi-Hanson metric), but the 
procedure becomes complicated.  

A few comments are in order here. The above decays can be seen directly
from the corresponding GLSMs by Vafa's arguments \cite{vafa}. Let us
briefly discuss this. Consider a $U(1)$ GLSM with three fields,
$\left(\phi_{-n},\phi_{k_1},\phi_{k_2}\right)$ with charges 
$\left(-n,k_1,k_2\right)$, and the D-term constraint is given by 
\be
-n|\phi_{-n}|^2+k_1|\phi_{k_1}|^2+k_2|\phi_{k_2}|^2+r=0
\ee
The $r\gg 0$ case has been considered in section 2. 
Consider the limit $r \ll 0$. In this case, the
D-term constraint implies that both $\phi_{k_1}$ and $\phi_{k_2}$ cannot be
simultaneously zero. Along with the $U(1)$ action, this implies that
the space is the weighted projective space $\BW\BP_{k_1,k_2}$. The 
$\phi_{-n}$ direction is a non-compact $O(-n)$ line bundle over 
this space, and hence the total space is the $O(-n)$ line bundle 
over $\BW\BP_{k_1,k_2}$, which is the complex space
\be
\left(\phi_{-n},\phi_{k_1},\phi_{k_2}\right) \simeq
\left(\lambda^{-n}\phi_{-n}, \lambda^{k_1}\phi_{k_1},
\lambda^{k_2}\phi_{k_2}\right)
\label{vafadec}
\ee
where both $\phi_{k_1}$ and $\phi_{k_2}$ cannot simultaneously vanish
and $\lambda \neq 0$ is a $C^*$ action. There are orbifold singularities
in this geometry. The geometry at $\left(\phi_{-n}= \phi_{k_1}=0\right)$ 
is locally $\BC^2/\BZ_{k_2}$ and that at 
$\left(\phi_{-n}= \phi_{k_2}=0\right)$ is locally $\BC^2/\BZ_{k_1}$ 
(as can be seen by appropriately choosing the parameter $\lambda$), and
these give the generic points to which the orbifold decays \cite{vafa}.
For $k_1=1$, this matches with the metric computation presented here.

Also note that, in order to consider tachyon condensation in various twisted
sectors, we need to modify the GLSM. In particular, for studying tachyon
condensation in the jth twisted sector of the orbifold $\BC^2/\BZ_{n(k)}$,
we consider the GLSM of three fields with $U(1)$ charges 
$\left(-n,j,jk\right)$. In this case, the sigma model metrics can be 
easily calculated as before. 

The form of the metric (without gauge fixing) in (\ref{metric1}) is 
interesting. Although the simultaneous angle identification in 
(\ref{iden1}) implies that the structure is not that of two separate
cones, nevertheless one is tempted to think that the height of the 
closed string tachyon potential would have a similar additive structure.
Let us see if we can substantiate this. Dabholkar \cite{dabhol} has
conjectured that the height of the closed string tachyon potential for
the orbifold of the form $\BC/\BZ_n$ is proportional to the deficit
angle of the cone and is given by    
\be
V_1(T) = 4\pi\left(1 - \frac{1}{n}\right)
\label{dabhform}
\ee
this conjecture was recently tested for certain orbifolds 
in bosonic closed string field theory \cite{oz} and upto $70$ 
percent agreement was found in certain transition processes between 
two different configurations. The height of the closed string tachyon
potential can also be computed from considerations of $N=2$ SCFT, 
where this height is given by the absolute value of the highest 
axial charge of the Ramond sector ground state \cite{dv}. This agrees
with (\ref{dabhform}) upon the substitution $n\to\frac{n}{2}$ \cite{dv}.
upto a possible normalization.

Consider the $U(1)$ GLSM for the orbifold $\BC^2/\BZ_n(k)$ for which the
metric has been derived in (\ref{metric1}) and apparently has an 
additive structure of two conical deficits. This was derived by taking
the $U(1)$ charges of the three fields to be $\left(-n,1,k\right)$. 
As we have mentioned, this in some sense assumes that the first twisted
sector of the orbifold is the most tachyonic (i.e has highest 
negative mass squared). If instead, the $j$th twisted sector contains
the most relevant tachyon, then the condensation of that tachyon is
described by an $U(1)$ GLSM with the charges of the fields being
$\left(-n,j,[jk]\right)$ where $[jk]$ is the integer part of $jk$.
In that case, the metric of (\ref{metric1})
should be modified to be
\be
ds^2=d\rho_1^2+\frac{\rho_1^2}{(n/j)^2}d{\tilde{\theta_1}}^2 +
d\rho_k^2 + \frac{\rho_k^2}{(n/[jk])^2}d{\tilde{\theta_2}}^2
\label{metric2}
\ee
with now the gauge invariant variables are ${\tilde\theta_1}=
\frac{n}{j}\theta_1+\theta_n$ and ${\tilde\theta_2}=
\frac{n}{kj}\theta_1+\theta_n$. The deficit angles of the two
apparent cones formed by ${\tilde\theta_1}$ and ${\tilde\theta_2}$
are given by $\left(1-\frac{j}{n}\right)$ and 
$\left(1-\frac{jk}{n}\right)$. This leads us to the following form
of the height of the tachyon potential for $\BC^2/\BZ_{n(k)}$
\be
V_2(T)=\left(1-\frac{2j}{n}\right)+\left(1-\frac{[2kj]}{n}\right)
\ee
Where j denotes the most relevant twisted sector, for which 
the tachyon mass is the maximum. It is not difficult to convince 
oneself that this formula is equivalent to the one in \cite{dv} 
(last equation of that paper). 
Essentially, the height of the closed string tachyon potential is 
proportional to the mass squared of the most relevant tachyon \cite{sin1}, 
and relates to the twisted sector with minimum R-charge. 
From a calculation of the sigma model
metric, this height, for closed string tachyon
condensation in the non-supersymmetric orbifold of the form
$\BC^3/\BZ_{n(p_1,p_2)}$ will be given by
\be
V_3(T)=\left(1-\frac{2j}{n}\right)+\left(1-\frac{[2p_1 j]}{n}\right) 
+\left(1-\frac{[2p_2 j]}{n}\right) 
\ee
which is again proportional to the mass of the most relevant tachyon
for the twisted sector $j$.

\section{The Coulomb Branch of the GLSM}

Now that we have studied the decay of non-supersymmetric orbifolds from
the point of view of the GLSM, we discuss another important
aspect of these orbifolds that can be studied using this model. As 
is well known, the classical ground state of the GLSM has both Higgs 
and Coulomb branches. The former is described as a solution to the 
D-term constraint of eq. (\ref{dterm}) and is obtained by assigning 
vevs to the scalars in the chiral multiplet $\phi_i$ (which is the scalar 
component of the chiral superfield $\Phi_i$ in eq. (\ref{glsmaction})). 
The scalars in the vector multiplet, $\sigma_a$ (the scalar component 
of the twisted chiral superfield $\Sigma_a$ in eq. (\ref{glsmaction}) 
remain massless in the process. The Coulomb branch solutions,
on the other hand are obtained as classical ground states of the theory
with non-zero vevs for some of the $\sigma_a$s, when some of the 
$\phi_i$ are zero. Typically, for space-time supersymmetric orbifolds, 
the Coulomb branch vacua are absent. They arise only in examples where 
the total $U(1)$ charge of the chiral fields are different from zero. 
These occur in the non-supersymmetric examples. In these latter examples,
the Higgs branch is the non-linear sigma model that has, as its target
space, the resolution of the orbifold singularity. 
   
As we have mentioned in the section (2.2), there is an apparent puzzle
here. In the resolution of non-supersymmetric orbifolds of the 
form $\BC^2/\BZ_n$, one finds that the Higgs branch, i.e the resolved 
space contains less number of D-brane charges
than that of the initial orbifold. For eg. from the toric data of the 
orbifold $\BC^2/\BZ_{8(3)}$ given in eq. (\ref{eightthree}), we can
calculate the charge matrix of the corresponding GLSM which is given
by the kernel of the matrix in eq. (\ref{eightthree}),
\be
{\cal Q}~=~
\pmatrix{ 3 & -8 & 0 & 1 \cr 1 & -3 & 1 & 0 \cr  }
\label{eightthreecharge}
\ee
Hence, this orbifold is described by the $U(1)^2$ GLSM, and there are
two Higgs branch vacua (corresponding to the blowing up of the two 
$\BP^1$s in the Hirzebruch-Jung continued fraction). However, 
the K-theory lattice of this orbifold is isomorphic to $\BZ^8$ 
(in general for the non-supersymmetric orbifold $\BC^2/\BZ_n$ the 
K-theory lattice will be isomorphic to $\BZ^n$) and thus there should 
be eight independent D-brane charges at the orbifold point, 
hence in this case, one is faced with the problem of the six 
missing D-brane charges.

This problem was solved in full generality in the elegant paper by
Martinec and Moore \cite{mm}. Among other things, These authors showed 
that by considering
the Coulomb branch of the GLSM (which is typically present in the 
non-supersymmetric orbifold examples) one is able to recover the full
set of independent D-brane charges. According to \cite{mm}, if the
non-supersymmetric orbifold of the form $\BC^2/\BZ_{n(k)}$ is described
by the Hirzebruch-Jung continued fraction $[a_1,\cdots,a_r]$, i.e the
resolution consists of blowing up $r$ $\BP^1$s, then the number of 
massive Coulomb branch vacua in the corresponding GLSM is $(n-r-1)$, which,
along with the $r$ Higgs branch solutions and the D-0 brane charge 
accounts for the expected total number $n$ of independent D-brane
charges. 

The analysis of \cite{mm} is general, and deals with turning on all the
Fayet-Iliopoulos parameters in a $U(1)^r$ GLSM that describes a singularity
of the form $\BC^2/\BZ_n$. However, it is clear from the analysis of
\cite{vafa} that since the resolution of singularity in a complex two-fold
orbifold is canonical, one can, in principle, turn on a single $U(1)$ at
a time, while studying the decay of the non-supersymmetric orbifolds of
$\BC^2$. This is what we had done while calculating the sigma model
metrics for the GLSMs for these orbifolds in the last subsection. 

Turning on a single F.I parameter at a time leads to an orbifold with a 
lower rank, and hence repeating this process, one can reach the end 
point of tachyon condensation, which is a supersymmetric configuration. 
From the continued fraction point of view, this (stepwise) resolution of
the singularity amounts to blowin up $\BP^1$s from one side. 
Note that before we reach a stable configuration, at each stage in the 
flow, we are dealing with a GLSM with $\sum Q_i \neq 0$. 
As is well known \cite{wittenphases},\cite{mp}, in a 
$U(1)$ GLSM where $\sum Q_i \neq 0$, the model, apart from the usual 
Higgs branch vacua, contains additional Coulomb branch vacua. This
happens when the $\sigma_a$ field gains a vev, hence giving mass to 
the $\phi_i$ fields, which results in a corrected twisted superpotential,
which in turn leads to a modification of the potential energy, and
the ground states corresponding to the modified potential energy are
the massive Coulomb branch vacua. In fact, it can be shown that for a $U(1)$
theory, the number of new Coulomb branch vacua is 
\cite{wittenphases},\cite{mp} 
\be
{\cal N}=|\sum Q_i| 
\label{cbv}
\ee
hence these vacua will arise in all the 
non-supersymmetric examples. We can use this information to calculate 
the number of Coulomb branch vacua for the resolution of the 
singularity $\BC^2/\BZ_n$. For example, the $\BC^2/\BZ_{n(1)}$
orbifold is described by a $U(1)$ GLSM with three chiral fields of
charges $(-n,1,1)$ and the number of Coulomb branch vacua in this case
is from eq. (\ref{cbv}) equal to $(n-2)$. For multiple $U(1)$ examples,
one can compute the number of these vacua by blowing up consecutive
$U(1)$s, i.e by turning on one Fayet-Iliopoulose parameter at a time.
We can track the decays by following Vafa's arguments \cite{vafa}. 

Let us start from the orbifold $\BC^2/\BZ_{n(k)}$ and follow its decay
using the arguments of \cite{vafa}. In order to do this, we set 
$k_1=1$ in eq. (\ref{vafadec}). Then, it is easy to see that 
we have the following decay process
\be
\BC^2/\BZ_{n(k)} \to \BC^2/\BZ_{k(2k-n)} + {\mbox{flat space}}
\label{vafadec1}
\ee
here we have assumed that $k < n, 2k > n$. Apriori, this need not be the
case, and  as we have discussed before, we will, in general, get
an orbifold of the form $\BC^2/\BZ_{k(pk-n)}$, However, this will not
affect the essence of our discussion, which will be applicable even
if $p\neq 2$. For the sake of simplicity, we assume that $2k > n$ in
what follows. Clearly, the orbifold $\BC^2/\BZ_{k(2k-n)}$ can have 
further singularities, and will decay using another tachyon direction, 
and the next step in the decay will be to the orbifold
$\BC^2/\BZ_{2k-n(3k-n)}$, etc. until finally a supersymmetric configuration
is reached.

There are two cases to consider here. First, the original 
non-supersymmetric singularity can decay into a singularity of the
form $\BC^2/\BZ_{l(1)}$ which finally decays into flat space, or, 
after a few steps of desingularisation (by turning on successive $U(1)$s)
it can reach a supersymmetric configuration of the form 
$\BC^2/\BZ_{l(-1)}$. \footnote{This would happen, for example, 
for orbifolds that have a continued fraction expansion of the form 
$[a_1,\cdots,a_m,2,\cdots,2]$.} Let us consider the first case. The original
non-supersymmetric orbifold will be in general a $U(1)^r$ GLSM. By blowing 
up $\BP^1$s corresponding to successive $U(1)$s, at each stage we 
obtain the number of Coulomb branch vacua for the GLSM (at that stage), 
and the counting goes as 
\begin{eqnarray}
n &-& k - 1\nonumber\\
&+&k - (2k-n) - 1 \nonumber \\
&+&~~~~(2k-n) - (3k - 2n) - 1 \nonumber\\
&+&~~~\cdots ~~~~~~~~~\cdots\nonumber\\
&+&~~~~~~~~~~~~~~~~~~~~~l-1-1
\label{cbv1}
\end{eqnarray}
Since the penultimate state (desingularisation of which leads to 
flat space) is of the form $\BC^2/\BZ_{l(1)}$, hence the last line
of eq. (\ref{cbv1}) is $(l-1-1)$, and the next step in the decay is
supersymmetric flat space. Also, there are $r$ lines in eq. (\ref{cbv1})
because of the initial $U(1)^r$ theory. On adding the lines 
of eq. (\ref{cbv1}), we see that the number of Coulomb branch vacua is
\be
{\cal N}=n-r-1
\ee
since there are $r$ factors of $1$ coming from each line, and the last
line contributing a final additive factor of unity. The second case is
similar. Suppose the continued fraction corresponding to the singularity
is of the form 
\be
\frac{n}{k}=\left[a_1,\cdots,a_{r-m},\underbrace{2,\cdots,2}_{m~ 
{\mbox{times}}}\right]
\ee
Then, since $[2,2,\cdots (m~{\mbox{times}})]=\frac{m+1}{m}$, 
it is clear that in eq. (\ref{cbv1}) the last line is of the form  
$m'-(m+1)-1$, where $m'$ is the rank of gauge group in the penultimate 
stage of the decay, the final stage being the orbifold denoted by
the continued fraction $[2,2,\cdots (m~{\mbox{times}})]$. Since we
have to turn on only $(r-m)$ Fayet-Iliopoulos parameters, there are
$(r-m)$ additive factors of unity, and hence the total number of 
Coulomb branch vacua is $n-(r-m)-(m+1) = (n-r-1)$. 

In addition to this, there will be the usual $r$ Higgs branch solutions,
and along with the D-0 brane charge, these account for the total number 
of independent D-brane charges, $n$. 

We emphasize that these results are well understood only for type 0 string
theory. For type II backgrounds, the chiral GSO projection can eliminate
some of the chiral ring generators, and a description of the resolved
space in terms of K\"ahler parameters only might not be possible. This case
needs to be understood better.

\section{Using Toric Methods}

In this section, we develop on the discussion in section (2.2) on the
use of methods from toric geometry to analyse non-supersymmetric orbifolds
of the form $\BC^2/\BZ_n$ and $\BC^3/\BZ_n$. 

\subsection{Toric Data for Orbifolds of $\BC^2$ and $\BC^3$}

We begin this subsection by pointing out a 
curious fact which will actually form the basis of our discussion in the
rest of the paper. In section (2.2), we have described the way to calculate 
the toric data of a given non-supersymmetric singularity from the brane 
probe approach, by using the orbifold projected gauge theory of the 
(type II) D-brane world volume. There is an interesting aspect in the
calculation of the toric data in this way. Namely, in the D-brane world
volume theory gives the toric data along with certain multiplicities 
in the same. In the calculation of the D-brane gauge theory as
in section (2.2), the number of fields $p_{\alpha}$ that appear in the
toric data are more than the number of $\BP^1$s 
which are needed to resolve the singularity, \footnote{The number of
fields $p_{\alpha}$ appearing in the probe analysis depends
on the dual matrix ${\vec T}$ of section (2.2) and it is not
possible to derive a general formula for the exact number of these fields,
and the procedure has to be repeated case by case. However, in most 
cases the number of these fields turns out to be more than the number of
vectors in the toric diagram of the resolution of the singularity. This 
happens for both complex two- and three-dimensional orbifolds} 
and more than one field $p_{\alpha}$ is seen to give rise to the same 
toric vector. This is not a triviality, and Hanany and collaborators have
shown that for supersymmetric $\BC^3$ orbifolds, the multiplicities of the
toric diagrams are related to dual gauge theories in the IR \cite{hantoric}.
These results are actually derived for type II branes, but this fact 
will not be important for our discussion below. 

Let us illustrate with the example of the orbifold
$\BC^2/\BZ_{5(3)}$. In this case, the number of fields $p_{\alpha}$ 
in the probe analysis is $12$, and the toric data is given by 
\be
{\cal T}~=~
\pmatrix{ 1 & 0 & -1 & -3 \cr 0 & 1 & 2 & 5 \cr {\bf 1}
 & {\bf 5} & {\bf 5} & {\bf 1}\cr }
\label{fivethreemult}
\ee
where in the last row of the matrix we have written the multiplicities
of the toric data points that arise in the brane probe analysis. 
Consider now the supersymmetric three-fold orbifold 
$\BC^3/\BZ_{5(1,3)}$ (in the notation of eq. (\ref{threefolds})
this is the orbifold with $p_1=1,p_2=3$). The toric data in this case
is obtained from $13$ fields, and is given by
\be
{\cal T}~=~
\pmatrix{ 1 & 0 & 0 & -1 & -3 \cr 
0 & 0 & 0 & 0 & -1 \cr
0 & 0 & 1 & 2 & 5 \cr 
{\bf 1} & {\bf 1} & {\bf 5} & {\bf 5} & {\bf 1}\cr }
\label{fivethreec3mult}
\ee
where again the last line denotes the multiplicities of the toric
data points. Note that by projecting the data of the orbifold 
$\BC^3/\BZ_{5(1,3)}$ on the height one plane corresponding to
the second row of this matrix, we get the toric data for the 
$\BC^2/\BZ_{5(3)}$ orbifold, along with an additional vector 
$(0,0)$. This is, in some sense, expected from \cite{asp}, 
but curiously, the multiplicities of the corresponding fields in the 
brane probe analysis are also the same, which apriori need not have 
been the case. Let us illustrate this with another example. Consider 
the non-supersymmetric orbifold of the form $\BC^2/\BZ_{8(5)}$. 
The toric data from the brane probe analysis (along with the 
multiplicities of the GLSM fields) is given by the matrix
\be
{\cal T}~=~
\pmatrix{ 1 & 0 & -1 & -3 & -5 \cr
0 & 1 & 2 & 5 & 8 \cr
{\bf 1} & {\bf 8} & {\bf 12} & {\bf 8} & {\bf 1}\cr }
\label{eightfivemult}
\ee
Comparing this with the data for the supersymmetric 
$\BC^3/\BZ_{8(2,5)}$ given by
\be
{\cal T}~=~
\pmatrix{ 1 & 0 & 0 & -1 & -2 & -3 & -5 \cr 
0 & 1 & 0 & 0 & 1 & -1 & -2 \cr
0 & 0 & 1 & 2 & 4 & 5 & 8 \cr 
{\bf 1} & {\bf 1} & {\bf 8} & {\bf 12} & {\bf 2} &
{\bf 8} & {\bf 1} \cr }
\label{eightfivec3mult}
\ee
we find that the multiplicities of the toric data points 
for the non-supersymmetric $\BC^2/\BZ_{8(5)}$
can be obtained from the supersymmetric $\BC^3/\BZ_{8(2,5)}$ (after 
appropriately projecting the data and 
removing certain additional points). What we are
actually doing is a {\it supersymmetric completion} of the 
non-supersymmetric two fold, into a supersymmetric three-fold, 
by adding an extra field in the
theory, and the supersymmetric three-fold orbifold seems to capture
much of the data for the non-supersymmetric two-fold.
 \footnote{In the cases of the $\BC^2$ orbifolds, as we 
have mentioned, the brane probe analysis often gives a non-minimal 
resolution corresponding to some marginal deformation being turned on. 
In these cases also the data for the $\BC^3$ orbifold gives the 
correct multiplicities of the corresponding lower rank 
$\BC^2$.} 

Although the above results were derived for the type II brane, 
they point towards a generic feature of $\BC^3$ orbifolds, namely, 
it seems that it might be possible to study certain aspects
of non-supersymmetric two-fold orbifolds from an analysis of three 
fold orbifolds, and vice versa. In the GLSM construction, when expanded 
in terms of component fields, the kinetic term of the chiral fields
of eq. (\ref{glsmaction}) contains a term 
$-2\sum_a Q_{i,a}^2|\sigma_a|^2|\phi_i|^2$. Starting from a higher
dimensional theory, if we give a vev to one of the $\phi_i$ fields (so
that in effect we get a lower dimensional theory with the remaining 
massless $\phi_i$ fields), we see that this acts like a mass term for 
the $\sigma_a$ field. Hence, there might be an interesting 
interpolation between the Higgs and Coulomb branches of two theories 
of different dimensions. 

As we have seen, the relationship between the multiplicities of the 
different dimensional orbifolds gives us a hint that we may think of 
higher dimensional orbifolds in terms of lower dimensional ones. 
In the next subsection, we try to make this connection more precise.

\subsection{Orbifolds of $\BC^3$ from Orbifolds of $\BC^2$}

A study of higher-dimensional orbifolds starting from lower
dimensional ones have already been initiated in the mathematics 
literature, and a simple prescription for the same has been presented
by Craw and Reid \cite{cr}. By using their results, it turns out 
that a complex three-fold
orbifold can in fact be described by a suitable combination of 
{\it three} complex-two folds. This is in turn related to the McKay
correspondence for three-fold orbifolds. We will not go into the details
of the mathematics here, and the interested reader is referred to 
\cite{cr}, but will quote the results which will be useful for our
purposes. It is best to illustrate this again with an 
example. Consider the orbifold $\BC^3/\BZ_{5(1,3)}$ for which the
toric data has been presented in eq. (\ref{fivethreec3mult}). The action
of the orbifolding group on the $\BC^3$ is given by
\be
\left(Z^1,Z^2,Z^3\right) \to \left(\omega Z^1,\omega Z^2,\omega^3 Z^3
\right)
\label{cr1}
\ee
where $\omega = \e^{\frac{2\pi i}{5}}$. It turns out that the resolution
of this orbifold can be described by considering in succession, the
action of $\BZ_5$ on three $\BC^2$ orbifolds, by taking two 
coordinates at a time in eq. (\ref{cr1}) and {\it patching}
the data together. In this case, the three orbifolds in question are
\begin{eqnarray}
\BC^2/\BZ_{5(1)}~&~&~: ~~~~~ {\mbox{action of}}~~~ \BZ_5 ~~~~ 
{\mbox{on}}~~~~ \left(Z^1,Z^2\right)
\nonumber\\
\BC^2/\BZ_{5(3)}~&~&~: ~~~~~ {\mbox{action of}}~~~~ \BZ_5 ~~~~ 
{\mbox{on}}~~~~ \left(Z^2,Z^3\right)
\nonumber\\
\BC^2/\BZ_{5(2)}~&~&~: ~~~~~ {\mbox{action of}}~~~~ \BZ_5 ~~~~ 
{\mbox{on}}~~~~ \left(Z^3,Z^1\right)
\label{cr2}
\end{eqnarray}
where in the last line, we have considered the action 
$\left(Z^3,Z^1\right) \to \left(\omega^3 Z^3, \omega Z_1\right)$ which
is equivalent to the one in (\ref{cr2}). \footnote{We could of course 
work with the action on the coordinates $\left(Z^1,Z^3\right)$ but for 
book keeping purposes, it would be more convenient to follow a cyclic 
notation as in \cite{cr}.} A combination of these $\BC^2$ orbifolds has 
to be done in accordance with the Hirzebruch-Jung continued fraction 
expansion, and these are:
\begin{eqnarray}
\BC^2/\BZ_{5(1)}&~&:~~~~ \frac{5}{1}~=~\left[5\right]\nonumber\\
\BC^2/\BZ_{5(3)}&~&:~~~~ \frac{5}{3}~=~\left[2,3\right]\nonumber\\ 
\BC^2/\BZ_{5(2)}&~&:~~~~ \frac{5}{2}~=~\left[3,2\right]
\label{cr3}
\end{eqnarray}
We now apply the prescription of 
Craw and Reid for the construction of the toric data for the 
resolution of $\BC^3/\BZ_{5(1,3)}$. Essentially, this amounts to 
combining the three $\BC^2$ orbifolds in (\ref{cr3}), by drawing 
them as three corners of a triangle (the so called ``junior simplex''), 
and drawing lines representing the terms in the continued fractions 
in (\ref{cr3}) from the 
corresponding vertices. The rule is that the strength of each line
is the value of the integer it represents in the continued fraction.
When two or more lines meet, the one with greater strength defeats
the one with lesser strength, but the strength of the former decreases
by one for each line it defeats. Further, lines meeting with 
equal strengths all die. This produces a triangulation whose 
associated toric variety precisely represents the resolution of the orbifold
$\BC^3/\BZ_{5(1,3)}$.
\footnote{In some cases we need to join certain interior points of 
our triangle.} We have illustrated this in figure (\ref{fig1}). 
\begin{figure}
\centering
\epsfxsize=4.5in
\hspace*{0in}\vspace*{.2in}
\epsffile{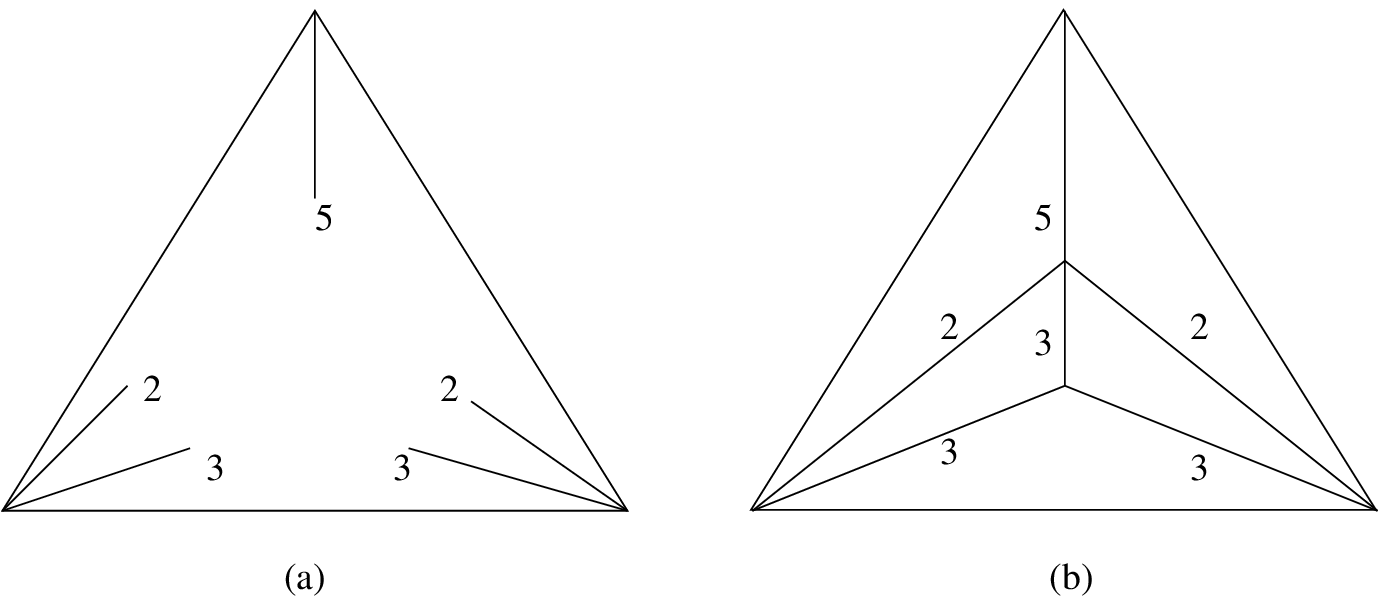}
\caption{\small The ``champions meet'' for the orbifold 
$\BC^3/\BZ_{5(1,3)}$.} 
\label{fig1}
\end{figure}
The three continued fractions of eq. (\ref{cr3}) are drawn at the 
three vertices representing the three $\BC^2$s in (a). In (b),
we show the combination according to the prescription just described, 
meet, which gives the resolution of the orbifold $\BC^3/\BZ_{5(1,3)}$. 

Now, let us try to understand the above from the point of view of
conformal field theory. Here, the construction according to the above
prescription should boil down to the determination of the chiral ring
of the $\BC^3$ orbifold starting from three $\BC^2$ orbifolds. For the 
latter, Martinec and Moore have shown that the convex region of the plot of 
the R-charge vectors $\left(\frac{j}{n},\frac{jp}{n}\right)$ of the twist 
operators of eq. (\ref{twist}) (forming the Newton boundary) 
can be used to determine the (twisted) 
chiral ring of the $\BC^2$ orbifolds \cite{mm}. For eg., for the orbifold 
$\BC^2/\BZ_{5(1)}$, there are four twisted sectors, with the R-charge 
vectors being$\left(\frac{1}{5},\frac{1}{5}\right), 
\left(\frac{2}{5},\frac{2}{5}\right), \left(\frac{3}{5},\frac{3}{5}\right),
\left(\frac{4}{5},\frac{4}{5}\right)$. A plot of these charges is shown
in fig. (\ref{fig2}) (a). From the convex region formed by these charges
(along with the volume forms for the two coordinate directions represented by
the two points on the two axes), we see that the chiral ring consists 
of a single generator in this case, which is of course the operator
with charge $\left(\frac{1}{5},\frac{1}{5}\right)$. Similarly, from the
figures (\ref{fig2}) (b) and (c), which shows the R-charges for the
twist operators for the orbifolds $\BC^2/\BZ_{5(3)}$ and 
$\BC^2/\BZ_{5(2)}$, the corresponding generators of the 
chiral rings can be determined. 
\begin{figure}
\centering
\epsfxsize=4.5in
\hspace*{0in}\vspace*{.2in}
\epsffile{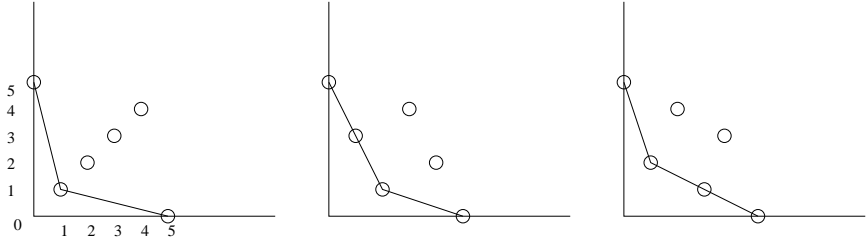}
\caption{\small R-charges of the generators of the chiral rings of the 
three $\BC^2$ orbifolds represented by the three vertices of 
fig. (\ref{fig1}). The labelling of the axes in the second
and third diagrams are as in the first one.}
\label{fig2}
\end{figure}
We summarise these generators :
\begin{eqnarray}
\BC^2/\BZ_{5(1)}&~&:~~~~ \left(\frac{1}{5},\frac{1}{5}\right) 
\nonumber\\
\BC^2/\BZ_{5(3)}&~&:~~~~ \left(\frac{1}{5},\frac{3}{5}\right),
\left(\frac{2}{5},\frac{1}{5}\right)\nonumber\\ 
\BC^2/\BZ_{5(2)}&~&:~~~~ \left(\frac{1}{5},\frac{2}{5}\right),
\left(\frac{3}{5},\frac{1}{5}\right)\nonumber\\ 
\label{cr4}
\end{eqnarray}
It is now clear how to combine the three $\BC^2$ orbifolds. This
reduces to appropriately combining the generators of the $\BC^2$
orbifolds to form those of the $\BC^3$ orbifold. For example, 
taking the generator $\left(\frac{3}{5},\frac{1}{5}\right)$ of 
$\BC^2/\BZ_{5(2)}$, we need to join it with the generator 
$\left(\frac{1}{5},\frac{3}{5}\right)$ of the orbifold 
$\BC^2/\BZ_{5(3)}$ for consistency of the R-charges corresponding
to the various coordinates of the $\BC^3/\BZ_{5(1,3)}$ orbifold, and
this is a generator of the chiral ring of the orbifold 
$\BC^3/\BZ_{5(1,3)}$, with R-charges 
$\left(\frac{1}{5},\frac{1}{5},\frac{3}{5}\right)$
There is another possibility of joining the R-charge vectors of the $\BC^2$
orbifolds in a consistent way, and these two points produce the 
combination of fig. (\ref{fig1}). This 
translates the construction of \cite{cr} to the language
of chiral rings. It was in some sense expected, given the product
structure of the conformal field theory, but note that in this 
construction, we crucially needed the fact that three different 
$\BC^2$ orbifolds are being glued together. This was the non-trivial
ingredient in the discussion. 

We have illustrated one of the simplest examples here. In general
there can be complications. For eg., for supersymmetric orbifolds of 
the form $\BC^3/\BZ_{n(p_1,p_2)}$, the situation is fundamentally 
different for the cases $p_1=p_2$ and $p_1\neq p_2$. Whereas in the
first case the topology of desingularisation does not allow for 
flop transitions, in the second case, there might be several phases
of the theory connected by flops \cite{muto}. However, in both cases
the supersymmetric $\BC^3$ orbifolds can be constructed out of their $\BC^2$ 
cousins, much in the same way as we have described in the previous
paragraph, and this construction holds for both type 0 and type II
theories. 
 
We now discuss how the above construction helps us to study the
decay of orbifolds.  

\subsection{Decay of Orbifolds of $\BC^3$}

Decay of orbifolds of $\BC^3$ can be studied in the same way as we
have discussed in section 3. Here, we will elaborate on this, and show
how for threefold orbifolds, the resulting decays can be seen from the
underlying two-folds, following the construction of the previous
subsection. As a warmup, it will be useful for us to first consider 
the decay of
certain two-dimensional examples. Let us begin with supersymmetric
$A_{n-1}$ orbifolds of the form $\BC^2/\BZ_{n(n-1)}$, whose toric
data is given by eq. (\ref{data1}). We can consider both type 0 and type II
strings here. As discussed in \cite{hkmm}, 
we can study the closed string CFT for these orbifold backgrounds, 
and its decays by perturbing the Lagrangian by marginal operators
(there are no tachyons in this case). In complete analogy, one can
analyse these decays by using Vafa's method of blowing up $\BP^1$s
in succession. Let us apply eq. (\ref{vafadec1}) to the supersymmetric
orbifold $\BC^2/\BZ_{4(3)}$ whose toric data is obtained from eq.
(\ref{data1}) as
{\small
\begin{eqnarray}
{\cal T}=
\pmatrix{
1&0&-1&-2&-3\cr
0&1&2&3&4\cr
}\label{data1a}\end{eqnarray}}
This theory has three marginal deformations, corresponding to the 
chiral operators in the three twisted sectors, and these are the
three interior vectors in the toric diagram (the toric fan is 
generated by the vectors $(1,0)$ and $(4,-3)$. According to eq. 
(\ref{vafadec1}), perturbing the Lagrangian by the chiral operator
corresponding to the first twisted sector yields the flow
\be
\BC^2/\BZ_{4(3)} \to {\mbox{flat space}}~~+~~\BC^2/\BZ_{3(2)}     
\label{susydec}
\ee
This can be seen from eq. (\ref{data1a}) by splitting the toric data
into two parts along the vector corresponding to the first marginal
deformation, i.e $(0,1)$ and gives
{\small
\begin{eqnarray}
\pmatrix{
1&0&-1&-2&-3\cr
0&1&2&3&4\cr
}\to\pmatrix{1&0\cr 0&1\cr}\oplus
\pmatrix{0&-1&-2&-3\cr 1&2&3&4\cr}
\label{data1b}\end{eqnarray}}
The first matrix gives the toric data for flat space, corresponding to
the first term of eq. (\ref{susydec}), the second matrix, can, by 
the $SL(2,\BZ)$ matrix $\pmatrix{2&1\cr -1&0\cr}$
be brought to the form $\pmatrix{1&0&-1&-2\cr 0&1&2&3\cr}$ which 
is the toric data for the orbifold $\BC^2/\BZ_{3(2)}$, the second
term of eq. (\ref{susydec}). The fact that $SL(2,\BZ)$ transformations
can be used to read off the toric data was already noted in
\cite{ts1}. More recently, \cite{mnp} has used the Smith normal 
form algorithm (which amounts to performing $GL(2,Z)$ transformations
on toric data matrices) in order to read off toric data for complex
three-fold orbifolds. 

The above decay can be studied by the $U(1)$ GLSM for 
the orbifold $\BC^2/\BZ_{4(3)}$. It has three fields with charges 
$\left(-4,1,3\right)$. For studying the decay corresponding to the 
marginal perturbation with the second twisted sector, we need to use
the GLSM with charges $\left(-4,2,2\right)$, and this gives the flow
of the orbifold to two other supersymmetric orbifolds of the form 
$\BC^2/\BZ_{2(1)}$. This can be equivalently constructed by 
splitting the toric data of eq. (\ref{data1a}) into two parts
along the vector $(2,-1)$ and using an appropriate $SL(2,\BZ)$ 
transformation on the second part. A similar procedure can be followed
for the non-supersymmetric $\BC^2$ orbifolds \cite{ts1}. 

Now we move on to the decay of $\BC^3$ orbifolds. As a first step, let us 
see how to study the decay of supersymmetric 
orbifolds of $\BC^3$, by the methods described above. In this case,
a straightforward applications of the methods of \cite{vafa}
yield the following flow pattern for the flow corresponding to the
turning on of the chiral operator of the first twisted sector 
(described by an $U(1)$ GLSM of four chiral
fields with charges $\left(-n,1,p_1,p_2\right)$)
\be
\BC^3/\BZ_{n(p_1,p_2)}\to~~{\mbox{flat space}}\oplus
\BC^3/\BZ_{p_1(2p_1-n,p_2)}\oplus
\BC^3/\BZ_{p_2(p_1,2p_2-n)}
\label{vafac3}
\ee
This is applicable again for both marginal and tachyonic perturbations.
Consider the supersymmetric orbifold $\BC^3/\BZ_{5(1,3)}$ whose 
champions meet has been illustrated in fig. (\ref{fig1}). There are   
four twisted sectors here, out of which the first and second sectors 
are marginal, the third and fourth being irrelevant. Hence only the 
first two twisted sectors appear in the toric diagram of the 
resolution of this orbifold. We can consider perturbing the worldsheet 
Lagrangian by any of these marginal operators, in analogy with the 
two-dimensional examples. We summarise the flow patterns in 
fig. (\ref{fig3}).
\begin{figure}
\centering
\epsfxsize=5.5in
\hspace*{0in}\vspace*{.2in}
\epsffile{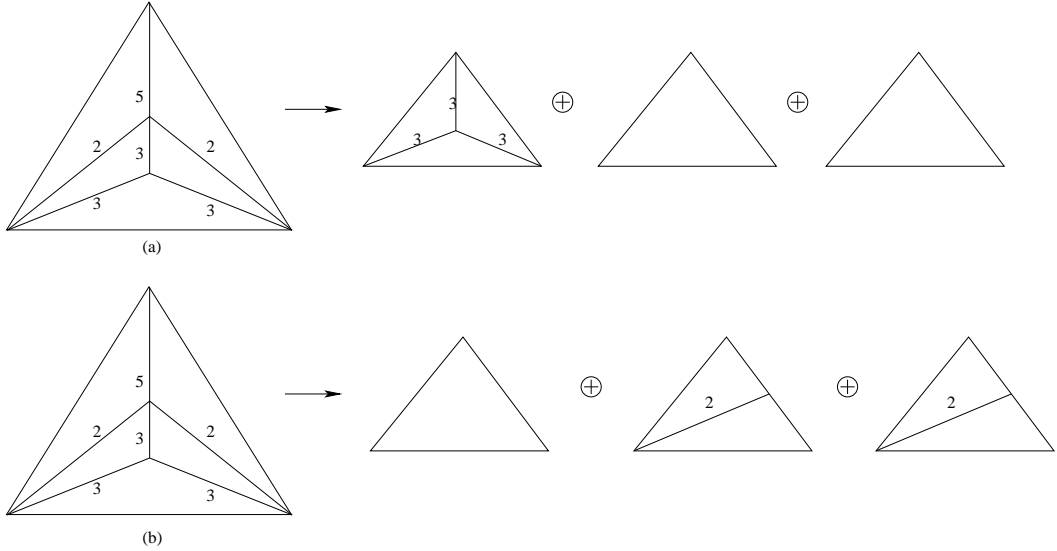}
\caption{\small Flow of the orbifold $\BC^3/\BZ_{5(1,3)}$ under
marginal deformations} 
\label{fig3}
\end{figure}
For the marginal deformation by the first twisted sector, eq. (\ref{vafac3}) 
gives the flow pattern
\be
\BC^3/\BZ_{5(1,3)}\to~~{\mbox{flat space}}~~\oplus~~
{\mbox{flat space}}~\oplus \BC^3/\BZ_{3(1,1)}  
\ee
This is shown in fig. (\ref{fig3})(a), where we have split the original 
triangle along the first twisted sector. Similarly, for deforming the
theory by the second twisted sector, we need to consider the $U(1)$ 
GLSM with four chiral fields of charges $\left(-5,2,2,1\right)$ and
in this case the flow is given by
\be
\BC^3/\BZ_{5(1,3)}\to~~{\mbox{flat space}}~~\oplus~~
\BC^3/\BZ_{2(1,0)}\oplus \BC^3/\BZ_{2(0,1)}
\ee
which is given by the diagram in fig. (\ref{fig3})(b). Interestingly,
the strength of the continued fractions are not modified in this 
example. This shows that the decay of supersymmetric $\BC^3$ orbifolds
can be studied from non-supersymmetric $\BC^2$ orbifolds.  

The example we just worked out was an orbifold of the form
$\BC^3/\BZ_{n(p,p)}$. It turns out that orbifolds of the form
$\BC^3/\BZ_{n(p_1,p_2)}$ with $p_1$ and $p_2$ distinct have a
different description, and our method is not very robust in this
case. In fact, while the former type of orbifolds do not contain
topologically distinct phases, the latter type does. For example, 
the orbifold $\BC^3/\BZ_{11(2,8)}$ contains topologically 
distinct phases related by flop transitions \cite{muto}, corresponding
to distinct ways of triangulating the toric diagram for the resolution
of this orbifold. This is reflected in the fact that in this case, 
a combination of the $\BC^2$ orbifolds 
does not provide an unique triangulation of the
toric diagram. This is evident from fig. (\ref{fig4}). The dotted
lines indicate in this example the extra lines that we need in order
to triangulate the diagram completely (which has $11$ triangles of
area $1$), and the solid lines are the Hirzebruch-Jung lines, i.e 
correspond to the continued fractions that appear in the $\BC^2$
orbifolds.
\begin{figure}
\centering
\epsfxsize=3.5in
\hspace*{0in}\vspace*{.2in}
\epsffile{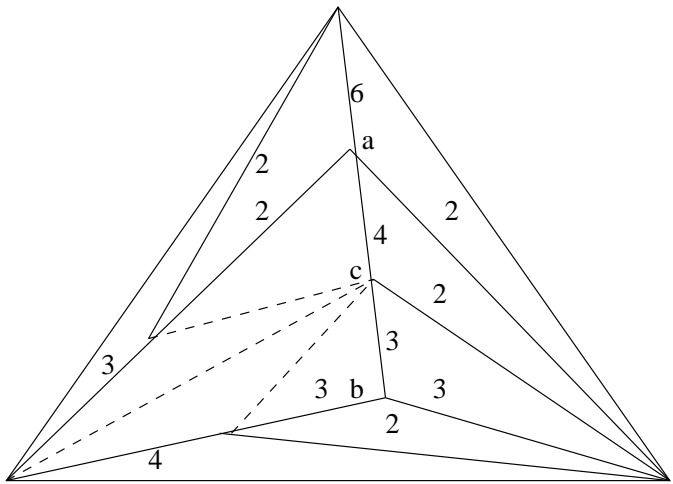}
\caption{\small Champions meet for the orbifold $\BC^3/\BZ_{11(2,8)}$ 
marked in solid lines. The strengths of the Hirzebruch-Jung continued 
fractions for the three $\BC^2$ orbifolds are marked with each solid
line, which we call the Hirzebruch-Jung lines.} 
\label{fig4}
\end{figure}
Now let us study flow patterns in this supersymmetric example. Consider
first turning on the marginal deformation corresponding to the first
twisted sector of the theory. This is the point
$a$ in fig. (\ref{fig4}), \footnote{The R-charges of the Hirzebruch-Jung
lines can be evaluated following the procedure of \cite{mm}, and hence
the twisted sectors corresponding to the different points in the 
triangulation can be obtained easily.} with R-charges 
$\left(\frac{1}{11},\frac{2}{11},\frac{8}{11}\right)$. 
From Vafa's construction, this will lead to the decay 
\be
\BC^3/\BZ_{11(2,8)}\to~{\mbox{flat space}}\oplus
\BC^3/\BZ_{2(1,0)}\oplus
\BC^3/\BZ_{8(2,5)}
\ee
This is seen from fig. (\ref{fig4}), with the Hirzebruch-Jung
integers that have been depicted. \footnote{We need to be careful 
about the action of the discrete group on the various coordinates. 
For example, the decomposition of the orbifold $\BC^3/\BZ_{8(2,5)}$
into two-fold orbifolds contains as an action
$\left(X<Y\right)\to\left(\omega^2 X,\omega^5 Y\right)$ where 
$\omega$ is the eighth root of unity. We need to interchange coordinates
in this case in order to cast this action to canonical form. One needs
to keep track of such changes carefully.} Similarly, we can study the
decay of this orbifold by turning on a marginal deformation corresponding
to the third twisted sector of the theory, with R-charges
$\left(\frac{3}{11},\frac{6}{11},\frac{2}{11}\right)$ which is
point $b$ in fig. (\ref{fig4}). Here, the decay products
are three orbifolds of $\BC^3$ with the quotienting groups being 
$\BZ_2, \BZ_3, \BZ_6$, and this decay also can be observed from the 
figure, as before. However, the decay of this orbifold by turning on
a marginal deformation corresponding to the second twisted sector
(point $c$ in fig. (\ref{fig4})) is difficult to analyse. This case
is complicated because it depends on the phase of the theory being looked
at, and the flow cannot be seen as easily as in the previous examples. 
In particular, we observe that it is difficult to analyse flows that
involve triangles which have as one or more edges a line that is not a
Hirzebruch-Jung line of any of the $\BC^2$ orbifolds. This case needs to be 
studied further. 

Finally, let us consider the flows of non-supersymmetric orbifolds 
of $\BC^3$, which has been dealt with recently in \cite{mnp}. In this
case, there is no analogue of the construction of \cite{cr} 
as in supersymmetric examples. 
The latter was a mathematical construction dependent on the
fact that the orbifold quotienting group is a finite subgroup of 
$SL(3,\BC)$, and those methods cannot be applied for non-supersymmetric
$\BC^3$ orbifolds. However, since we have translated the results to the
language of chiral rings earlier in this section, let us try to cast 
the problem in terms of the generators of the chiral rings. We would like to 
emphasise that this is just a book-keeping technique for these examples,
and does not lead to any mathematical construction analogous to \cite{cr}.  

Again there are two distinct cases to consider here. We will
consider one example each of orbifolds of the form 
$\BC^3/\BZ_{n(p_1,p_2)}$, with $p_1=p_2$ and $p_1\neq p_2$. 
\footnote{Actually when the orbifold action is
of the form $\left(Z^1,Z^2,Z^3\right)\to
\left(\omega^{p_1}Z^1,\omega^{p_2}Z^2,\omega^{p_3}Z^3\right)$, it
is enough to consider the cases when two of the $p_i$s are equal
or they are all distinct.}
Let us first consider type 0 string theory on $\BC^3/\BZ_{13(5,5)}$. From the
discussion of the previous subsection, we can construct the chiral
ring of this orbifold (the generators of the ring are 
three elements of R-charges
$\left(\frac{1}{13},\frac{5}{13}\frac{5}{13}\right),
\left(\frac{3}{13},\frac{2}{13}\frac{2}{13}\right),
\left(\frac{8}{13},\frac{1}{13}\frac{1}{13}\right)$) by using
the chiral ring generators of the orbifolds $\BC^2/\BZ_{13(5)}$,
$\BC^2/\BZ_{13(1)}$ and $\BC^2/\BZ_{13(8)}$. These are the tachyonic
(and possibly irrelevant) operators of the world sheet CFT. 
Further, since we can evaluate the Hirzebruch-Jung integers for these 
three $\BC^2$ orbifolds, 
we can draw a diagram analogous to the 
supersymmetric examples, with the elements of the chiral ring
and the corresponding Hirzebruch-Jung integers. 
This is shown in fig. (\ref{fig5}).
\begin{figure}
\centering
\epsfxsize=3.5in
\hspace*{0in}\vspace*{.2in}
\epsffile{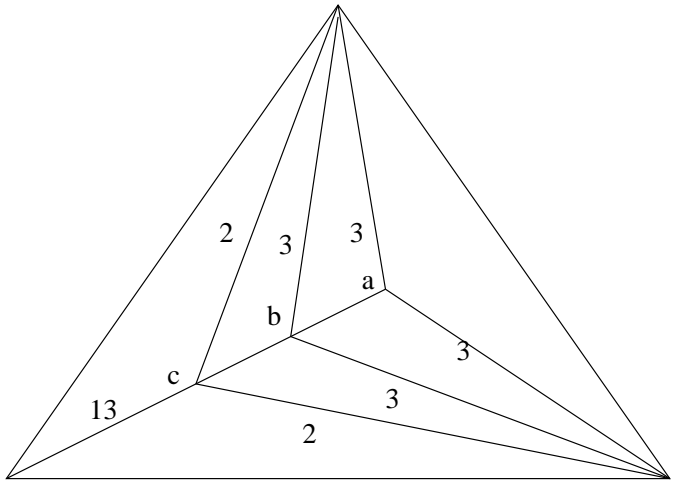}
\caption{\small An analogue of the champions meet for the 
orbifold $\BC^3/\BZ_{13(5,5)}$. The Hirzebruch-Jung integers 
label each line.} 
\label{fig5}
\end{figure}
Let us now consider the decay of this orbifold by the relevant 
operator corresponding to the first twisted sector of this orbifold,
which is point $a$ in fig. (\ref{fig5}). By our previous analysis,
this would correspond to the flow
\be
\BC^3/\BZ_{13(5,5)} \to {\mbox{flat space}} \oplus
\BC^3/\BZ_{5(2,0)} \oplus \BC^3/\BZ_{5(2,0)}
\ee
This flow can be seen from fig. (\ref{fig5}), by considering the
Hirzebruch-Jung numbers of the corresponding $\BC^2$ orbifolds. 
Similarly, the decay of this orbifold by the tachyon of point 
$b$ (which appears in the third
twisted sector) leads to the flow into three $\BC^3$ orbifolds with
the ranks of the orbifolding group being $\BZ_3, \BZ_2, \BZ_2$, 
and this can also be seen from the Hirzebruch-Jung numbers of  
fig. (\ref{fig5}). However, we need to be 
careful here. The Hirzebruch-Jung number of the line that is split in 
the process of the decay (in this case the line with strength $13$) 
should be ignored while calculating the action of the orbifolds in the 
decay process, and the action of the resulting orbifolds should be fixed 
by the remaining lines. \footnote{This seems to be a
generic feature in studying these flows. It is presumably
related to a renormalisation of the R-charges under tachyon condensation
as in \cite{mnp}, and we leave a detailed study of this for the future.}
It is not difficult to convince oneself that
this uniquely fixes the orbifold action (we will make this clearer in
the next example).

Finally, let us consider the type 0 orbifold $\BC^3/\BZ_{13(2,5)}$ which
has been studied in details in \cite{mnp}. The Hirzebruch-Jung 
continued fraction gives the four generators of the chiral ring in 
this example, and they are as shown in fig. (\ref{fig6}).  
\begin{figure}
\centering
\epsfxsize=3.5in
\hspace*{0in}\vspace*{.2in}
\epsffile{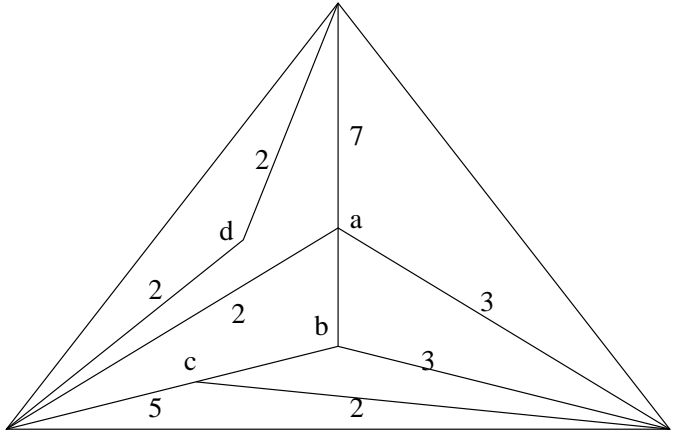}
\caption{\small An analogue of the champions meet for the 
orbifold $\BC^3/\BZ_{13(2,5)}$. The interior points are the
generators of the chiral ring.} 
\label{fig6}
\end{figure}
Consider now the condensation of the tachyon corresponding to the
first twisted sector (point $a$ in fig. (\ref{fig6}), and
called $T_1$ in \cite{mnp}). From
the figure, we see that this give three $\BC^3$ spaces, one of which 
is flat space, one is the terminal singularity $\BC^3/\BZ_{2(1,1)}$
and the third is the supersymmetric orbifold $\BC^3/\BZ_{5(2,2)}$. 
This also follows from eq. (\ref{vafac3}).
\footnote{The last orbifold is realised as follows: from our previous 
discussion, we ignore the line coming from the top vertex. Out of the 
coordinates $\left(Z^1,Z^2,Z^3\right)$, the Hirzebruch-Jung
numbers $[2,3]$ fixes the action on two of the coordinates
($Z^3$ and $Z^1$ as the orbifold $\BC^2/\BZ_{5(3)}$. 
The other number similarly fixes the action on $(Z^2,Z^3)$ as the
orbifold $\BC^2/\BZ_{5(1)}$. This fixes the orbifold to be
$\BC^3/\BZ_{5(2,2)}$.} 

The tachyon of point $b$ in fig. (\ref{fig6}) comes from the
third twisted sector of the theory. Thus, in order to study its 
condensation, we need to consider the corresponding GLSM of four
fields with charges $\left(-13,3,6,2\right)$. It is clear that the 
condensation of this tachyon leads to three $\BC^3$ orbifolds with
quotienting groups $\BZ_3, \BZ_6$ and $\BZ_2$. This can also be
seen from figure (\ref{fig6}) by considering the Hirzebruch-Jung
numbers. The details of the orbifold action can be worked out 
and the results match with \cite{mnp}. Condensation of the tachyon
of the eighth twisted sector (marked point c) is however difficult
to follow using our method. 

\section{Conclusions}

In this paper, we have studied in details certain issues regarding
localized tachyon condensation on orbifolds of $\BC^2$ and $\BC^3$. 
We have used the GLSM to understand various aspects of the same. 
We have seen that a computation of the background metric for the 
GLSM effectively captures much of the physics of closed string
tachyon condensation, in lines with \cite{vafa}. The GLSM also
gives a nice way of understanding the number of Coulomb branch
vacua of non-supersymmetric backgrounds. Further, we have shown 
that the decay of $\BC^3$ orbifolds can be studied in terms of 
orbifolds of $\BC^2$. 

There are various issues still to be studied. We have
mostly considered flows in type 0 string theory, and there are 
several subtleties that might arise for type II strings, for eg., some
of the generators of the chiral ring may be projected out for these
\cite{mm}, and hence the resolution of the singularity cannot be
described entirely by K\"ahler deformations, as we have mentioned
before. It will be interesting to understand the interpretation
of our results of section 5 in this case. 

Secondly, our analysis shows that it is possible to study localized
closed string tachyon condensation on orbifolds of the form 
$\BC^3/\BZ_{n(p_1,p_2)}$ starting from two-fold orbifolds of the
form $\BC^2/\BZ_{n(k)}$. A better understanding of this is 
can probably be obtained from the point of view of the GLSM. 
Specifically, one can ask if the moduli space of GLSMs corresponding 
to orbifolds of $\BC^3$ contain, in certain (probably unphysical) 
limits the orbifolds of $\BC^2$. We have already seen indications of 
this in section 5. It would be interesting to explore this issue further. 

Conversely, we have also seen that toric geometry of $\BC^3$ 
orbifolds contains information about orbifolds of $\BC^2$, and
in most cases there are additional points that need to be removed
from the former in order to obtain the latter. It might be possible
to study this further from the inverse toric algorithm proposed
in \cite{invtoric}. Also, as shown in \cite{ts1}, D-brane gauge
theories on orbifolds of the form $\BC^2/\BZ_{n(k)}$ appear to contain
various ``phases.'' However, since the brane probe analysis presented
there was valid in the substringy regime, no statement could be made
about any duality between these phases, and it was essentially a
classical statement about the moduli space of D-brane world volume
gauge theories. It might be possible to study such phases
by supersymmetric completion of non-supersymmetric two-fold orbifolds 
to supersymmetric three-folds. 

Finally, the issue of obtaining lower dimensional orbifolds starting
from the higher dimensional ones might be studied by the boundary
state formalism. This would help in unifying the 
various issues discussed above.\\

\noindent
{\bf Acknowledgements}

We would like to sincerely thank the High Energy Theory Group of
Harvard University for their hospitality during the course of
this work. Its a pleasure to thank Justin David, Suresh Govindarajan, 
Ami Hanany, Dileep Jatkar, Shiraz Minwalla, Koushik Ray, 
Tadashi Takayanagi and Cumrun Vafa for useful discussions. Thanks 
are due to Utpal Chattopadhyay for computer related help.

\end{document}